\documentclass[journal]{IEEEtran}[11pt]
\ifCLASSINFOpdf
\usepackage[pdftex]{graphicx}
\graphicspath{{../pdf/}{../jpeg/}}
\DeclareGraphicsExtensions{.pdf,.jpeg,.png}
\else
\usepackage[dvips]{graphicx}
\graphicspath{{../eps/}}
\DeclareGraphicsExtensions{.eps}
\fi\usepackage{graphicx}

\usepackage{subfloat}
 
\usepackage{graphics}
\usepackage{epsfig}
\usepackage{epstopdf}
\usepackage{stfloats}
\usepackage[cmex10]{amsmath}
\usepackage{algorithmic}
\usepackage{array}
\usepackage{mdwmath}
\usepackage{mdwtab}
\usepackage{graphicx}
\usepackage{subfigure}
\usepackage{color}
\usepackage{amsfonts,amssymb}
\usepackage{hyperref} %

\usepackage{multirow}
\usepackage{diagbox}
\usepackage{caption}
\usepackage{makecell}
\usepackage{lineno}
\usepackage{mathrsfs}

\usepackage{booktabs}
\usepackage{makecell}
\usepackage{tabu}
\usepackage{multirow}
\usepackage{multicol}
\usepackage{multirow}
\usepackage{float}
\usepackage{makecell}
\usepackage{booktabs}
\usepackage{amssymb, bm}
\usepackage{bbding}
\usepackage[linesnumbered,ruled]{algorithm2e}
\usepackage{amsfonts,amsthm,array}
\usepackage[ruled]{algorithm2e}

\usepackage{threeparttable}

\begin{document}
	

\title{Joint 3D Trajectory Design and Resource Allocation for Secure Dual-UAV-aided Underlay Systems}

\author{Hongjiang~Lei, ~\IEEEmembership{Senior Member,~IEEE,}
	Xiaqiu~Wu,
	Ki-Hong~Park, ~\IEEEmembership{Senior Member,~IEEE,}\\
	Gaojie~Chen, ~\IEEEmembership{Senior Member,~IEEE}
	and
	Gaofeng~Pan, ~\IEEEmembership{Senior Member,~IEEE}
	\thanks{This work was supported by the National Key Research and Development Program of China under Grant 2024YFC3306801, National Natural Science Foundation of China under Grant 62571045, and Natural Science Foundation of Chongqing under Grant (cstc2024ycjh-bgzxm003, CSTB2025NSCQ-LZX0053).	(Corresponding author: \textit{Hongjiang Lei}.)}
	\thanks{Hongjiang~Lei and Xiaqiu~Wu are with the School of Communications and Information Engineering, Chongqing University of Posts and Telecommunications, Chongqing 400065, China (e-mail: leihj@cqupt.edu.cn, cquptwxq@163.com).}
	\thanks{Ki-Hong~Park is with the CEMSE Division, King Abdullah University of Science and Technology (KAUST), Thuwal 23955-6900, Saudi Arabia (e-mail: kihong.park@kaust.edu.sa).}
	\thanks{Gaojie~Chen is with the School of Flexible Electronics (SoFE), Sun Yat-sen University, Shenzhen, Guangdong 518107, China (e-mail: gaojie.chen@ieee.org).}
	\thanks{Gaofeng~Pan is with the School of Cyberspace Science and Technology, Beijing Institute of Technology, Beijing 100081, China (e-mail: gfpan@bit.edu.cn).}
}

\maketitle

\begin{abstract}

Uncrewed aerial vehicles (UAVs) are increasingly being employed for data collection tasks, thanks to their high mobility and easy deployment, acting as aerial platforms to collect data from ground devices (GDs). This study considers a secure underlay data collection system assisted by dual UAVs and focuses on the joint design of the UAVs' three-dimensional (3D) flight paths, the power of the jamming UAV, the power of GDs, and the scheduling of the underlay GDs in the context of an aerial eavesdropper. The highly coupled objective function and non-convex constraints make the formulated problem more complicated to solve.
{ We first utilize an approximate lower bound on the expected {spectral efficiency} to streamline the solution process. 
The average secrecy spectral efficiency (ASSE) is maximized by jointly designing the 3D trajectory of the UAVs, the transmit power of GDs, and the user scheduling. } The optimization problem is decomposed into four subproblems using block coordinate descent, with each of them into manageable convex optimization tasks by incorporating slack variables and employing successive convex approximation methods. The numerical results validate the effectiveness of our proposed approach, demonstrating that the design of UAV 3D trajectories remarkably improves the ASSE of the considered system.
	
\end{abstract}
\begin{IEEEkeywords}
	3D trajectory design, 
	physical layer security, 
	aerial eavesdropping, 
	cognitive radio networks,
	power control, 
	data collection.
	
\end{IEEEkeywords}

\section{Introduction}
\label{sec:introduction}

\subsection{Background and Related Works}
\label{Background}

Uncrewed aerial vehicles (UAVs) have undergone remarkable development, evolving from traditional military roles to become essential assets in a variety  of civilian and commercial applications. Today, UAVs play a pivotal role in various scenarios, including modern agriculture, disaster management, surveillance, and aerial imaging \cite{WuQ2021JSAC}. The swift advancements, particularly in domains such as autonomy, payload capacity, and flight duration, have significantly expanded their functionality and applicability across various industries \cite{LiB2019IOT, ZengY2019P}.
However, the rapid development also presents a host of new opportunities and challenges, highlighting the necessity for ongoing research in critical areas such as regulatory frameworks, security schemes, and the integration of UAVs with existing infrastructure. A fundamental advantage of UAV-aided systems is the high-probability line-of-sight (LoS) link between the UAV and ground devices (GDs). 
Consequently, designing the optimal trajectory of UAVs is essential for enhancing the efficiency of UAV communication networks and has emerged as a significant challenge in the evolution of UAV-assisted systems \cite{WuQ2019WC1}. As the deployment of UAVs continues to increase, addressing these trajectory planning issues will be crucial for unlocking their full potential and ensuring their seamless integration into various applications. 

Physical layer security (PLS) utilizes the distinctive characteristics of wireless channels to establish robust defenses against the risks of eavesdropping. 
For instance, \textit{in the scenarios with terrestrial eavesdroppers}, the authors in Ref. \cite{CuiM2018TVT} focused on optimizing both the trajectory of the UAV and its transmit power to maximize the worst-case average secrecy rate (ASR) of UAV-aided communication systems with multiple location-uncertain eavesdroppers.
Ref. \cite{LiZ2019CL} explored a multifaceted optimization problem that encompassed the UAV's flight path, transmit power, and user associations, and an iterative algorithm was proposed to maximize the secrecy rate while simultaneously ensuring fairness among the GDs involved.
Moreover, the authors in Ref. \cite{NaZ2022IOT} examined the secrecy performance of cooperative aerial Internet of Things (IoT) systems. They proposed a novel algorithm to maximize the minimum ASR by jointly optimizing the UAV's trajectory and resource allocation.
{ Refs. \cite{AAl2023IWCMC} and \cite{AAl2025OJCS} employ deep reinforcement learning to improve the security and energy efficiency of UAV systems. In particular, the utility function consisted of secrecy rate and energy utilization efficiency and was jointly maximized by optimizing the UAV's trajectory, power allocation, and harvesting energy.
Furthermore, the use of UAVs as airborne jammers has proven effective in enhancing the security of wireless communication systems, largely due to their ability to adjust their trajectories flexibly \cite{KimM2021TVT}–\cite{ZhangR2021TWC}.  
These studies confirm that deploying aerial jammers can significantly improve communication secrecy performance, even when the eavesdropper's location is imperfectly known. However, the strategies proposed in the aforementioned works are specifically designed for fixed ground-based eavesdroppers and are not directly applicable to scenarios involving mobile aerial eavesdroppers.}

\textit{
	Compared to traditional terrestrial fixed eavesdroppers, aerial platforms present far greater security risks due to their mobility, adaptability, and ability to access direct signal paths—a threat that has been emphasized in many works \cite{LiB2019WC, MH2019WC, WuQ2019WC}.}
For instance, the authors in \cite{LuW2022TCOM} considered a UAV-assisted mobile edge computing (MEC) system with an aerial eavesdropper, and the non-orthogonal multiple access (NOMA) technology was utilized to deal with the interference among the GDs. The ergodic computation capacity was maximized by optimizing various critical parameters, including user scheduling, transmit power, UAV flight paths, and the allocation of computational resources.
Ref. \cite{DingY2024TWC} proposed a collaborative strategy for an aerial communication and computation MEC system with an aerial eavesdropper and focused on minimizing energy consumption by optimizing UAV trajectory, transmit power of GDs, time, and offloading computation allocation. 
Additionally, Ref. \cite{ChengT2023SJ} explored the secrecy performance of aerial communication networks augmented by intelligent reflecting surfaces (IRS) and focused on optimizing transmit power, beamforming techniques, and the flight trajectories of UAVs to mitigate the risks associated with aerial eavesdropping effectively.
Moreover, the authors of \cite{MichailidisET2024TCOM} proposed a security-aware computation offloading framework designed explicitly for UAV-enabled IoT networks, which face threats from both terrestrial and aerial eavesdroppers. 
The secrecy outage probability was analyzed, and the minimum secure computation efficiency was maximized by jointly designing power and task allocation, time slot scheduling, and phase shifts.
In a related vein, Ref. \cite{LeiH2024MASS} focused on reducing the time-averaged computational costs in NOMA-based aerial MEC systems by jointly optimizing UAV trajectories, transmit power from GDs, and computational frequency.
Ref. \cite{LeiH2024TCCNFair} and \cite{LeiH2025TAES} sought to maximize the fair secrecy goodput in UAV-aided communication systems consisting of multiple aerial base stations, GDs, and a flying eavesdropper. Their approach involved coordinated designs for both UAVs' 2D/3D trajectories and transmit power, aiming to achieve enhanced security of the considered system and fairness among GDs. 
Ref. \cite{AlhabobAA2025TCOM} considered an integrated sensing and communication (ISAC) system with multiple aerial eavesdroppers with unknown locations, and a maximum likelihood-based scheme was proposed to estimate the eavesdroppers' channels by utilizing a long short-term memory deep neural network. 
Ref. \cite{GaoX2025TVT} investigated the secrecy performance of a space-air-ground (SAG) network with an airship eavesdropper. Considering both with and without a no-fly zone, the trajectory and transmit power of the aerial relay were jointly optimized to maximize the secrecy energy efficiency of the SAG system. 
Ref. \cite{WangQ2025TMC} investigated the secrecy performance of a terrestrial communication system with a ground jammer (GJ) and multiple aerial eavesdroppers. 
The secrecy capacity of the considered system was maximized through jointly designing the trajectory and transmit power of the GJ.

Cognitive radio networks (CRNs) is recognized as a pioneering solution to the spectrum scarcity issues in UAV-assisted communication systems, leveraging dynamic spectrum access methodologies \cite{SaleemY2015JNCA}. 
In the underlay mode, cognitive users (CUs) can utilize their frequency bands when the interference from CUs to primary users (PUs) is limited to a given threshold, known as the interference temperature (IT).
Ref. \cite{HuangY2019TCOM} investigated the cognitive UAV communication systems in both stationary and mobile scenarios. 
Their goal was to enhance the CU's achievable data rate by integrating the UAV's 3D position and trajectory, as well as the transmit power of the ground transmitter, subject to constraints imposed by altitude, power limitations, and IT thresholds.
In Ref. \cite{WangZ2021CC}, a UAV worked as a relay and forwarded a signal from the cognitive transmitter (CT) to the cognitive receiver (CR). The throughput of the cognitive system was maximized by jointly optimizing the transmit power at CT and the UAV's 3D trajectory.
For instance, \textit{to investigate the secrecy performance of UAV-aided CRNs,}
Ref. \cite{GaoY2021IWCMC} addressed challenges associated with jamming in aerial CRNs and formulated an optimization problem to enhance throughput in sensitive areas by strategically designing the UAV's trajectory and transmit power to counter hostile jamming attacks effectively.
Additionally, Ref. \cite{LeiH2024TVT} concentrated on the joint design of the UAV's flight path and transmit power, explicitly targeting the optimization of ASR against colluding terrestrial eavesdroppers in aerial CRNs.
It is noteworthy that the works mentioned above model the air-ground link as an LoS link. The optimal 3D position/trajectory mainly depends on both IT constraints and the performance of the CR result.
In practical scenarios, the probability of an air-to-ground link being LoS is a function of the elevation angle, the relative position between the UAV and GDs, as well as the distribution of building density and height \cite{HouraniAAL2014WCL}. 
The results in \cite{DuoB2020TVT3D}-\cite{Lei2024TCCN} verified that \textit{there is a trade-off between the distance and the elevation angle in the probabilistic LoS (PLoS) model. {The PLoS channel model introduces the randomness of channel states, resulting in the system rate appearing as a random function, which significantly increases the complexity of joint optimization of UAV trajectories and resources.}}
Ref. \cite{JiangY2021CC} investigated maximizing the secrecy rate of a UAV-enhanced CRN, taking into account not only the UAV's spatial trajectory but also its velocity and acceleration. Their formulated optimization problem respected constraints related to UAV positioning requirements, speed limits, and the IT constraint across all PUs.
Ref. \cite{NguyenX2021TVT} introduced the use of a UAV as an adversarial jammer, generating artificial noise to suppress terrestrial eavesdropping. Considering the conditions of both perfect and imperfect eavesdropper locations, the CUs' ASR was optimized through adjustments of transmit power and the UAV's 3D trajectory.

\begin{table*}
	\centering
	
	\caption{Comparison with the state-of-the-art on UAV systems.}
	\label{table1}
	\begin{threeparttable}
		\resizebox{0.66\textwidth}{!}
		{
			\begin{tabular}{c|c|c|c|c|c|c|c|c} 
				\Xhline{1.2pt}
				\textbf{Reference} & \textbf{\makecell[c]{Number of \\ UAVs}} & \textbf{Role of UAVs} & \textbf{\makecell[c]{Channel\\ Model}} & \textbf{PLS} &\textbf{\makecell[c]{3D \\Trajectory}} & \textbf{\makecell[c]{Aerial \\Eavesdropper}} & \textbf{\makecell[c]{User \\Scheduling}} & \textbf{\makecell[c]{CRN}}\\		
				\hline
				\cite{CuiM2018TVT}	  & Single & Transmitter & LoS & $\checkmark$ &  & & & \\
				\hline
				\cite{LiZ2019CL}	  & Single & Transmitter & LoS & $\checkmark$ &  & & $\checkmark$ & \\
				\hline
				\cite{NaZ2022IOT}	  & Single & Relay & LoS & $\checkmark$ &  & & $\checkmark$ & \\
				\hline
				\cite{KimM2021TVT}	  & Single & Jammer & LoS & $\checkmark$ &  & & & \\
				\hline
				\cite{ZhouY2018TVT}	  & Single & Jammer & PLoS & $\checkmark$ & $\checkmark$ & & & \\
				\hline
				\cite{LiA2019WCL}	  & Single & Jammer & LoS & $\checkmark$ &  & & & \\
				\hline
				\cite{LiuZ2024TVT}	  & Single & Jammer & LoS & $\checkmark$ &  & & & \\
				\hline
				\cite{LeiH2023IoT}	  & Double & Transmitter and Jammer & LoS & $\checkmark$ &  & & $\checkmark$ & \\
				\hline
				\cite{ZhangR2021TWC}	  & Double & Receiver and Jammer & LoS & $\checkmark$ &  & & $\checkmark$ & \\
				\hline
				\cite{DingY2024TWC}	  & Single & Transmitter & LoS & $\checkmark$ &  & $\checkmark$ & $\checkmark$ & \\
				\hline
				\cite{ChengT2023SJ}	  & Single & Receiver & LoS, NLoS & $\checkmark$ & $\checkmark$ & $\checkmark$ & & \\
				\hline
				\cite{MichailidisET2024TCOM} & Single & Transmitter & LoS & $\checkmark$ &  & $\checkmark$ & & \\
				\hline
				\cite{LeiH2024MASS}	  & Single & Receiver & PLoS & $\checkmark$ & $\checkmark$ & $\checkmark$ & $\checkmark$ & \\
				\hline			
				\cite{LeiH2024TCCNFair} & Double & Transmitter and Jammer & LoS, PLoS & $\checkmark$ &  & $\checkmark$ & $\checkmark$ & \\
				\hline
				\cite{LeiH2025TAES}	  & Double & Transmitter and Jammer & LoS, PLoS & $\checkmark$ & $\checkmark$ & $\checkmark$ & $\checkmark$ & \\
				\hline
				\cite{AlhabobAA2025TCOM} & & Eavesdroppers & LoS & $\checkmark$ &  & $\checkmark$ & & \\
				\hline
				\cite{GaoX2025TVT}	  & Single & Jammer & LoS & $\checkmark$ &  & $\checkmark$ & & \\
				\hline
				\cite{WangQ2025TMC}	  & & Eavesdroppers & PLoS & $\checkmark$ &  & $\checkmark$ & $\checkmark$ & \\
				\hline
				\cite{HuangY2019TCOM} & Single & Transmitter & LoS & &  & & $\checkmark$ & $\checkmark$ \\
				\hline
				\cite{WangZ2021CC}	  & Single & Relay & LoS & &  & & & $\checkmark$ \\
				\hline
				\cite{GaoY2021IWCMC} & Single & Receiver & LoS & $\checkmark$ & $\checkmark$ & & &$\checkmark$ \\
				\hline
				\cite{LeiH2024TVT}	  & Single & Transmitter & LoS & $\checkmark$ &  & & & $\checkmark$ \\
				\hline
				\cite{JiangY2021CC}	  & Single & Transmitter & PLoS & $\checkmark$ & $\checkmark$ & & $\checkmark$ & $\checkmark$ \\
				\hline
				\cite{NguyenX2021TVT} & Single & Jammer & LoS & & $\checkmark$ & & $\checkmark$ & $\checkmark$ \\
				\hline
				Our work & Double & Receiver and Jammer & LoS, PLoS & $\checkmark$ & $\checkmark$ & $\checkmark$ & $\checkmark$ & $\checkmark$ \\
				\Xhline{1.2pt}
		\end{tabular}}
	\end{threeparttable}
\end{table*}

\textbf{\textit{Summarization and Discussions:}}
Table \ref{table1} summarizes the differences among the discussed work on UAV communication systems from various perspectives. 
The motivations for each column are specifically described as follows:  
(1) Number of UAVs: Existing studies are categorized based on whether they involve a single UAV or multiple UAVs.  
(2) Role of UAVs: A UAV may serve as a transmitter, a jammer, a relay, a receiver, or an eavesdropping node.  
(3) Channel Model: A LoS channel implies that the communication rate depends solely on distance, while a PLOS channel reflects real-world scenarios with building obstacles by offering a trade-off between elevation angle and distance.  
(4) PLS: Incorporating PLS entails greater complexity due to the presence of eavesdroppers, making UAV trajectory optimization more challenging, yet it provides practical guidance for secure communication.  
(5) 3D Trajectory: Compared to 2D trajectories, 3D trajectories can better leverage the advantages of UAVs, though they introduce higher complexity and pose greater research challenges.  
(6) Aerial Eavesdroppers: Mobile aerial eavesdroppers present more significant security challenges compared to fixed ground-based eavesdroppers.  
(7) User Scheduling: Effective user scheduling helps avoid inter-user interference and maximizes the benefits of UAV mobility in terms of spectral efficiency (SE), fairness, security, etc.   
(8) CRN: The integration of CRN can enhance SE, albeit at the cost of increased system complexity.

According to Table \ref{table1}, extensive research has been conducted on the secrecy performance of UAVs. However, the existing studies still present the following limitations.
1) In research involving a single UAV, the air-to-ground channels are generally modeled as LoS links. Most studies focus on ground-based stationary eavesdroppers. Only Refs. \cite{DingY2024TWC} and \cite{MichailidisET2024TCOM} considered aerial eavesdropping scenarios, but only in a 2D and non-CRN framework. Although studies such as \cite{ChengT2023SJ} and \cite{LeiH2024MASS} incorporated 3D scenarios, they remain limited to non-CRN systems.  
2) In studies utilizing multiple UAVs, deploying a friendly jammer UAV is widely recognized as an effective method to enhance system security performance. However, these works were all based on non-CRN systems. Moreover, only \cite{LeiH2025TAES} considered the PLoS channel model, while Refs. \cite{LeiH2023IoT}, \cite{ZhangR2021TWC}, and \cite{LeiH2024TCCNFair} were confined to 2D and LoS links and Refs. \cite{LeiH2023IoT} and  \cite{ZhangR2021TWC} assumed with terrestrial eavesdroppers. 
\textit{Compared to single-UAV systems, dual-UAV systems require simultaneous optimization of 3D trajectories for both UAVs, leading to a significant increase in the number of optimization variables. Moreover, these variables are interdependent, exacerbating the non-convexity of the problem.}

\color{black}
\subsection{Motivation and Contributions}
\label{Motivation}

The works discussed above demonstrate that aerial eavesdropping poses a significant threat to wireless communication security, while the use of a cooperative aerial jammer shows promising potential to enhance anti-eavesdropping performance. However, to the best of our knowledge, no prior work has addressed aerial eavesdropping in dual-UAV-assisted underlay CRN. To bridge this gap, we propose a novel framework that incorporates an aerial eavesdropper with imperfect location information and a ground-to-air (G2A) PLoS channel model. We jointly optimize the 3D trajectories of both UAVs, user scheduling, and the transmit power of ground devices (GDs) in a data-collection scenario to maximize the average secrecy SE (ASSE) of the system. The main contributions of this work are summarized as follows. 
\begin{enumerate}		
	\item We consider a data collection system where multiple GDs transmit confidential information to an aerial base station (BS) assisted by a jammer UAV, in the presence of a mobile eavesdropper attempting to intercept the communications. Our goal is to maximize the ASSE through the joint optimization of the two UAVs' 3D trajectories and transmit power, as well as the scheduling of the GDs. 
	This work considers more practical and comprehensive factors, including the practical PLoS channel, which leads to coupling among variables in the optimization, since the UAV trajectories involve both elevation angle and distance. Moreover, the interference experienced by primary users includes that from both GDs and the jammer UAV, across both fading and PLoS channels, making the problem highly complex to handle. Furthermore, we incorporate energy consumption for the 3D movement of UAVs while ensuring secure communication, taking into account an aerial mobile eavesdropper with imperfect channel state information (CSI) and CRN principles.	
	\item To tackle this non-convex problem, we utilize a lower bound approximation for the ASSE and decompose the original problem into four subproblems using block coordinate descent (BCD). By applying successive convex approximation (SCA), each subproblem is transformed into a tractable convex form. 
	Simulation results demonstrate that the proposed method outperforms benchmark schemes, confirming that optimizing 3D trajectories significantly improves the ASSE of the system.
\end{enumerate}

\color{black}

\subsection{Organization}

The rest of this work is organized as follows. The system model and problem formulation are provided in Section \ref{SystemModel} and \ref{Problem}, respectively. A two-step alternating algorithm is proposed in Section  \ref{ProposedAlgorithm1} to solve the problem. Simulation results are demonstrated in Section \ref{Simulation}. Finally, Section \ref{Conclusions} concludes this work. 
TABLE \ref{table02} lists the notations and symbols utilized in this work.
\textit{Notations}: It should be noted that boldface lowercase letters, boldface uppercase letters, and regular letters denote vectors, matrices, and scalars, respectively.

\begin{table}
	\centering
	\caption{Notation and Symbols.}
	\begin{threeparttable}
		\resizebox{0.45\textwidth}{!}
		{
		\begin{tabular}{ c | c  }
			\Xhline{1.2pt}
			{Notation}  & {Description}				\\
			\hline
			${\mathbf{q}}_U^{\mathrm I},{\mathbf{q}}_J^{\mathrm I},{\mathbf{q}}_E^{\mathrm I} $     & Initial location of $U$, $J$, and $E$           \\
			\hline
			${\mathbf{q}}_U^{\mathrm F},{\mathbf{q}}_J^{\mathrm F},{\mathbf{q}}_E^{\mathrm F}$     & Final location of $U$, $J$, and $E$  \\
			\hline
			${H_{\max }}$/${H_{\min }}$          &Maximum/minimum flight altitude  		\\
			\hline
			${V_{xy }}$/${V}_{z }$    	   & Maximum horizontal/vertical speed of $U$ and $J$		 \\	
			\hline
			${P_k^{\max }}$/${P_J^{\max }}$			& Maximum transmit power of $D_k$/$J$	\\
			\hline
			${P_k^{\mathrm {ave }}}$/${P_J^{\mathrm {ave }}}$		& Average transmit power of $D_k$/$J$	\\
			\hline
			${P_{{U,\mathrm {ave }}}^{{\mathrm{hor}}}}$/${P_{{U,\mathrm {ave }}}^{{\mathrm{ver}}}}$    & Average horizontal/vertical propulsion power of $U$  		\\
			\hline
			${P_{{J,\mathrm {ave }}}^{{\mathrm{hor}}}}$/${P_{{J,\mathrm {ave }}}^{{\mathrm{ver}}}}$    &  Average horizontal/vertical propulsion power of $J$  		\\
			\hline
			$\sigma^2$					& Noise power 			\\
			\hline
			${\alpha_{{\mathrm L}}},{\alpha_{{\mathrm N}}}$		&  Path loss exponents for LoS and NLoS links	\\		
			\hline
			${\mu}$					        	& Additional signal attenuation factor 				\\
			\hline
			${a},{b}$	                               & Environmental coefficients   \\
			\hline
			${\delta _t}$						& Time slot length 							\\
			\hline							
			${\varepsilon}$                   &  Algorithm convergence precision   \\
			\Xhline{1.2pt}
		\end{tabular}}
	\end{threeparttable}
\label{table02}
\end{table}

\section{System Model}
\label{SystemModel}

\begin{figure}[t]
	\centering		
	\includegraphics[width = 0.4 \textwidth]{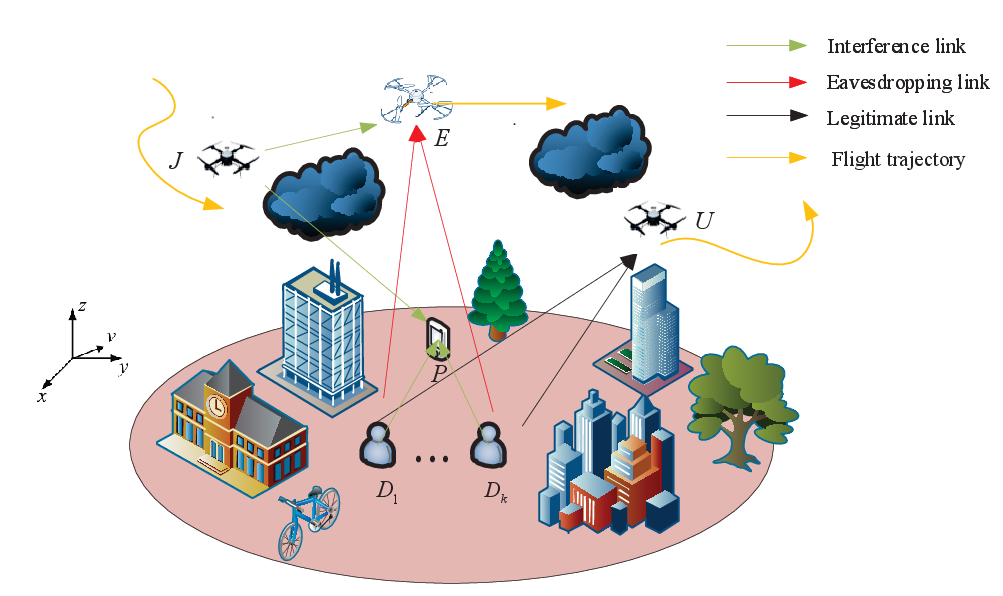}
	\caption{UAV-assisted secure data collection system consists of an aerial cognitive receiver (${U}$), a jamming UAV (${J}$), a potential eavesdropper (${E}$), a PU (${P}$), and multiple {fixed} GDs (${D_k}$).}
	\label{fig1model}
\end{figure}

As depicted in Fig. \ref{fig1model}, we investigate a dual UAV-assisted secure data collection system wherein a legitimate aerial cognitive receiver UAV (designated as $U$) collects confidential information from $K$ {fixed} GDs (represented as $D_k$ for $k = 1, 2, \ldots, K$). An additional UAV (referred to as $E$) operates as a potential eavesdropper, flying from an initial position $\mathbf{q}_E^{\mathrm{I}}$ to a final position $\mathbf{q}_E^{\mathrm{F}}$ during its flight \cite{DingY2024TWC}\footnote{
	{
		In this work, we consider the scenarios with a potential aerial eavesdropper (E), which is a common assumption in many works, like Refs. \cite{LeiH2023IoT}, \cite{LeiH2024TCCNFair}. 
		Specifically, $E$ performs some information-gathering tasks along a fixed trajectory and is ``curious" about the transmitted messages intended for GDs. 
		Moreover, in some special scenarios, some malicious eavesdropping UAVs may also be disguised as UAVs flying along a fixed trajectory to complete an open mission to collect some secret information.
	}
}.
To bolster communication security, a friendly jamming UAV, labeled $J$, generates artificial noise (AN) to suppress $E$, like \cite{KimM2021TVT}-\cite{LiuZ2024TVT}. 
It is assumed that the AN signals produced by $J$ take the form of Gaussian pseudo-random sequences or deterministic waveforms that mimic the structure of the desired signals \cite{XinH2016TVT}, \cite{LvL2019TIFS}, which allows $U$ to cancel the jamming signals perfectly. 
{Similarly, to conserve spectrum resources, it is assumed that the considered system operates in underlay mode \cite{WangZ2021CC}-\cite{NguyenX2021TVT}. 
Explicitly, the interference at a terrestrial primary user (PU), $P$, should be limited to a given threshold to guarantee reliable communication of $P$.}
All devices are assumed to be equipped with a single antenna. 
For simplicity, we discretize the flight duration $T$ into $N$ slots, such that the time index is represented as $n \in \mathcal{N} \triangleq \{1, 2, \ldots, N\}$ and each slot has a duration of $\delta_t = \frac{T}{N}$. We assume $N$ is sufficiently large so that the UAV's position can be approximated as constant within each slot \cite{LeiH2023IoT}. 
Without loss of generality, a Cartesian coordinate system is employed to denote the positions of all nodes. In the $n$-th slot, the horizontal coordinates of $U$, $E$, and $J$ are represented as $\mathbf{q}_U\left[ n \right] = [x_U\left[ n \right], y_U\left[ n \right]]^T$, $\mathbf{q}_E\left[ n \right] = [x_E\left[ n \right], y_E\left[ n \right]]^T$, and $\mathbf{q}_J\left[ n \right] = [x_J\left[ n \right], y_J\left[ n \right]]^T$, respectively. Their vertical positions are denoted by $z_U\left[ n \right]$, $z_E\left[ n \right]$, and $z_J\left[ n \right]$, respectively. The horizontal coordinates of  $D_k$ and $P$ are given by $\mathbf{w}_{k} = [x_k, y_k]^T$ and $\mathbf{w}_{P} = [x_P, y_P]^T$, respectively. 

{The air-to-air (A2A) communication link between UAVs is modeled as an LoS channel and the channel coefficient between $J$ and $E$ in the $n$-th slot is expressed as \cite{LeiH2023IoT}, \cite{ZhangR2021TWC}
\begin{align}
	h_{JE}\left[ n \right]=\frac{\rho_0}{\left\|\mathbf{q}_J\left[ n \right]-\mathbf{q}_E\left[ n \right]\right\|^2+|z_J\left[ n \right] - z_E\left[ n \right]|^2},
\end{align}
where $\rho_0$ represents the channel gain at a unit reference distance in an LoS environment. }

Like \cite{LeiH2023IoT} and \cite{Lei2024TCCN}, to enable all the GDs to be served fairly during flight time, 
a binary variable $ a_k\left[ n \right] $ is defined to denote the user scheduling such that $ a_k\left[ n \right] = 1 $ indicates that $D_k$ is active and communicating with $ U $ during the $n$-th slot. Otherwise, $ a_k\left[ n \right] = 0 $. It is assumed that only one GD can transmit data to $ U $ in each slot\footnote{
	In this work, we consider a scenario where GDs are geographically dispersed. The assumption of serving one GD per time slot not only simplifies system design but also ensures fair access among users.
	Furthermore, our framework can be extended to closely located GDs through orthogonal access schemes (e.g., TDMA), while the proposed power optimization also provides valuable insights for NOMA-aided scenarios.
	}. Therefore, we have
\begin{subequations}
	\begin{align}
		\sum_{k=1}^K a_k\left[ n \right] \leq 1, &\forall n, \label{diaodu1} \\  
		a_k\left[ n \right] \in\{0,1\}, &\forall k, n. \label{diaodu2}
	\end{align}
\end{subequations}	

It is assumed that the G2A link adheres to a PLoS channel model and the probability of an LoS connection during the $n$-th slot is expressed as \cite{HouraniAAL2014WCL} 
\begin{align}
	P_{kX}^{{\mathrm L}}\left[ n \right]=\frac{1}{1+a e^{-b\left(\theta_{kX}\left[ n \right]-a\right)}}, X \in \left\{ {U,E} \right\},
\end{align}
where 
$\theta_{kX}\left[ n \right]=\frac{180}{\pi} \arctan\left(\frac{z_X\left[ n \right]}{\left\|\mathbf{q}_{X}\left[ n \right]-\mathbf{w}_k\right\|}\right)$ is the elevation angle between $D_k$ and $X$, $a > 0$ and $b > 0$ are constants specified by the actual environment. Correspondingly, the non-line-of-sight (NLoS) probability is obtained as $P_{k X}^{{\mathrm N}}\left[ n \right] =1 - P_{k X}^{{\mathrm L}}\left[ n \right]$. 
Based on the LoS or NLoS state, the channel coefficient between $D_k$ and $X$ in the $n$-th slot is expressed as $h_{kX}^{{\mathrm L}}\left[ n \right]=\rho_0 d_{kX}^{-\alpha_{{\mathrm L}}}\left[ n \right]$ or $h_{kX}^{{\mathrm N}}\left[ n \right]=\mu \rho_0 d_{kX}^{-\alpha_{{\mathrm N}}}\left[ n \right]$, respectively, 
where 
$\mu$ denotes the extra signal attenuation factor encountered in an NLoS environment. The path loss exponents for both LoS and NLoS situations are represented by $\alpha_{\mathrm{L}}$ and $\alpha_{\mathrm{N}}$, respectively. It is generally accepted that the relationship $\alpha_{\mathrm{L}} \leq \alpha_{\mathrm{N}}$ holds true \cite{MengA2022IoT}.	
Consequently, the expected {SE} from $D_k$ to $U$ at the $n$-th slot is obtained as  \cite{MengA2022IoT}
\begin{align}\label{expectedRkU}
	\bar{R}_{kU}\left[ n \right] = P_{kU}^{{\mathrm L}}\left[ n \right] R_{kU}^{{\mathrm L}}\left[ n \right] + P_{kU}^{{\mathrm N}}\left[ n \right] R_{kU}^{{\mathrm N}}\left[ n \right],
\end{align}
where 
$R_{kU}^{{\mathrm L}}\left[ n \right] = \log_2\left(1 + \frac{P_k\left[ n \right]h_{kU}^{{\mathrm L}}\left[ n \right]}{\sigma^2}\right)$, 
$R_{kU}^{{\mathrm N}}\left[ n \right] = \log_2\left(1 + \frac{P_k\left[ n \right]h_{kU}^{{\mathrm N}}\left[ n \right]}{\sigma^2}\right)$, 
$P_k\left[ n \right]$ refers to the transmit power of $D_k$, 
and $\sigma^2$ indicates the variance of the additive white Gaussian noise (AWGN). 
Similarly, the expected {SE} at $E$ in the $n$-th slot is expressed as \cite{AAl2023IWCMC}, \cite{AAl2025OJCS}
\begin{align}\label{expectedRkE}
	\bar{R}_{kE}\left[ n \right] = P_{kE}^{{\mathrm L}}\left[ n \right] R_{kE}^{{\mathrm L}}\left[ n \right] + P_{kE}^{{\mathrm N}}\left[ n \right] R_{kE}^{{\mathrm N}}\left[ n \right],
\end{align}
where 
$R_{kE}^{{\mathrm L}}\left[ n \right] = \log_2\left(1 + \frac{P_k\left[ n \right]h_{kE}^{{\mathrm L}}\left[ n \right]}{P_J\left[ n \right]h_{JE}\left[ n \right]+\sigma^2}\right)$, 
$R_{kE}^{{\mathrm N}}\left[ n \right] = \log_2\left(1 + \frac{P_k\left[ n \right]h_{kE}^{{\mathrm N}}\left[ n \right]}{P_J\left[ n \right]h_{JE}\left[ n \right]+\sigma^2}\right)$,
and 
$ P_J\left[ n \right] $ denotes the transmit power of $ J $.

In this work, we consider more practical scenario where the location of $E$ is imperfect for GDs and $J$, which results in $h_{k E}$ and $h_{J E}$ are inaccurate. 
The bounded error model is utilized \cite{LvL2019TIFS}\footnote{
	The results of this work can be easily extended to the scenarios wherein Gaussian CSI error model is utilized. 
	Specifically, the position of $E$ is expressed as ${{\bf{Q}}_{\mathrm{E}}} = {\widehat {\bf{Q}}_{\mathrm{E}}} + \Delta {\bf q}$, where 
	${\widehat {\bf{Q}}_{\mathrm{E}}}$ denotes the estimated position of $E$ and
	$\Delta {\bf q}$ denotes the error vector. 	
	For the Gaussian error model, there is $ {\Delta {\bf q}} $ is a bi-variate Gaussian. 
	The general method is to approximate ${d_{i,{\mathrm{E}}}}\left[ n \right]$ as its expectation ${\mathbb{E}\left[ {{d_{i,{\mathrm{E}}}}\left[ n \right]} \right]}$ \cite{DanQ2025TCCN}. 
}. 
In particular, 
$\hat{\mathbf{q}}_E \in \mathbb{R}^{2 \times 1}$ and $r_E$ are utilized to indicate the circle center and radius of the area wherein $E$ is located, respectively. It is assumed that $r_E$ is smaller than the distance between GDs and $E$, i.e. $\left\|\hat{\mathbf{q}}_E\left[ n \right]-{\mathbf{w}}_k\right\| \geq r_E$  \cite{LeiH2023IoT}, \cite{ZhangR2021TWC}. According to the triangle inequality, the distance between $D_k$ and $E$ is expressed by the lower bound as
\begin{align}
	\left\|{\mathbf{q}}_E\left[ n \right]-\mathbf{w}_k\right\| & \geqslant\left\|{\hat{\mathbf{q}}}_E\left[ n \right]-{\mathbf{w}}_k\right\|-\left\|\hat{\mathbf{q}}_E\left[ n \right]-{\mathbf{q}}_E\left[ n \right]\right\| \nonumber \\
	& \geqslant\left\|\hat{\mathbf{q}}_E\left[ n \right]-{\mathbf{w}}_k\right\|-r_E.
\end{align}
Similarly, the distance between $J$ and $E$ is expressed by
the upper bound as
\begin{align}
	\left\|{\mathbf{q}}_E\left[ n \right]-\mathbf{q}_J\left[ n \right]\right\| & \leqslant\left\|\hat{\mathbf{q}}_E\left[ n \right]-{\mathbf{q}}_J\left[ n \right]\right\|+\left\|\hat{\mathbf{q}}_E\left[ n \right]-{\mathbf{q}}_E\left[ n \right]\right\|  \nonumber \\
	& \leqslant\left\|\hat{\mathbf{q}}_E\left[ n \right]-{\mathbf{q}}_J\left[ n \right]\right\|+r_E.
\end{align}
Then, the bounds of $h_{k E}$ and $h_{J E}$ are obtained as
\begin{align}
	h_{kE}^{{\mathrm L}, {\mathrm{ub}}}\left[ n \right] = \frac{\rho_0}{\left(\left(\left\|\hat{\mathbf{q}}_E\left[ n \right]-{\mathbf{w}}_k\right\|-r_E\right)^2+z_E^2\left[ n \right]\right)^{\alpha_{{\mathrm L}}/2} },
\end{align}
\begin{align}
	h_{kE}^{{\mathrm N}, {\mathrm{ub}}}\left[ n \right] = \frac{\mu\rho_0}{\left(\left(\left\|\hat{\mathbf{q}}_E\left[ n \right]-{\mathbf{w}}_k\right\|-r_E\right)^2+z_E^2\left[ n \right]\right)^{\alpha_{{\mathrm N}}/2} },
\end{align}
and 
\begin{align}
	h_{JE}^ {\mathrm{lb}}\left[ n \right]=\frac{\rho_0}{\left(\left\|\mathbf{q}_J\left[ n \right]-\hat{\mathbf{q}}_E\left[ n \right]\right\| + r_E\right)^2+|z_J\left[ n \right] - z_E\left[ n \right]|^2}.
\end{align}
Furthermore, $\theta_{kE}^{\mathrm{ub}}\left[ n \right]=\frac{180}{\pi} \arctan\left(\frac{z_E\left[ n \right]}{\left(\left\|\hat{\mathbf{q}}_{E}\left[ n \right]-\mathbf{w}_k\right\| - r_E\right)}\right)$ is the upper bound of elevation angle between $D_k$ and $E$. Therefore, the upper bound probability of LoS link between $D_k$ and $E$ is obtained as
$P_{kE}^{{\mathrm L}, {\mathrm{ub}}}\left[ n \right]=\frac{1}{1+a e^{-b\left(\theta_{kE}^{ub}\left[ n \right]-a\right)}}$. 
So, the upper bound of expected {SE} at $E$ in the $n$-th slot is determined as 
\begin{align}\label{expectedRkE}
	\bar{R}_{kE}^{\mathrm{ub}}\left[ n \right] = P_{kE}^{{\mathrm L},{\mathrm{ub}}}\left[ n \right] R_{kE}^{{\mathrm L},{\mathrm{ub}}}\left[ n \right] + \left(1 - P_{kE}^{{\mathrm L},{\mathrm{ub}}}\left[ n \right] \right) R_{kE}^{{\mathrm N},{\mathrm{ub}}}\left[ n \right],
\end{align}
where 
$R_{kE}^{{\mathrm L},{\mathrm{ub}}}\left[ n \right] = \log_2\left(1 + \frac{P_k\left[ n \right]h_{kE}^{{\mathrm L},{\mathrm{ub}}}\left[ n \right]}{P_J\left[ n \right]h_{JE}^ {\mathrm{lb}}\left[ n \right]+\sigma^2}\right)$ 
and 
$R_{kE}^{{\mathrm N},{\mathrm{ub}}}\left[ n \right] = \log_2\left(1 + \frac{P_k\left[ n \right]h_{kE}^{{\mathrm N},{\mathrm{ub}}}\left[ n \right]}{P_J\left[ n \right]h_{JE}^{lb}\left[ n \right]+\sigma^2}\right)$. 
The lower bound of the secrecy SE (SSE) at $D_k$ is expressed as
\begin{equation}
	{\bar{R}_{\sec}}^k\left[ n \right]=\left[\bar{R}_{kU}\left[ n \right]-\bar{R}_{kE}^{{\mathrm{ub}}}\left[ n \right]\right]^{+} \label{securerate1},
\end{equation}
where $[x]^{+} \triangleq \max (x, 0)$.	

The propulsion power consumption of UAV $i$ in the horizontal and vertical directions are expressed as \cite{ZengY2019TWC}
\begin{align}\label{speedxy}
	&{P_{{i}}^{{\mathrm{hor}}}}\left[ n \right] = {P_0\left(1+\frac{3\left\|\mathbf{v}_{xy,i}\left[ n \right]\right\|^2}{U_{\mathrm {tip }}^2}\right)} + {\frac{1}{2} d_0 \rho s A\left\|\mathbf{v}_{xy,i}\left[ n \right]\right\|^3}  \nonumber\\
	& + {P_1\left(\sqrt{1+\frac{\left\|\mathbf{v}_{xy,i}\left[ n \right]\right\|^4}{4 v_0^4}}-\frac{\left\|\mathbf{v}_{xy,i}\left[ n \right]\right\|^2}{2 v_0^2}\right)^{1/2}}, \forall n
\end{align}
and 
\begin{align}\label{speedz}
	{P_{{i}}^{{\mathrm{ver}}}}\left[ n \right] = {W_i} v_{z,i}\left[ n \right], \quad \forall v_{z,i}\left[ n \right]>0, \forall n,
\end{align}
where 
$i \in \left\{ {U,J} \right\}$, 
$\mathbf{v}_{xy,i}\left[ n \right] = \frac{1}{\delta_t} \left( \mathbf{q}_{i}\left[ n + 1 \right] - \mathbf{q}_{i}[n] \right)$
represents the flying speed of UAV $i$ during the hovering state \cite{ZhanC2019WCL}, \cite{DuoB2020TVFD}, 
$P_0$ and $P_1$ represent the inherent power of the blade surface and the power induced by the rotor, respectively, 
$U_{\mathrm{tip}}$ signifies the tip speed of the rotor blade, 
$d_0$ and $\rho$ indicate the body resistance ratio and air density, respectively, 
$v_0$ denotes the average induced velocity of the rotor, 
$s$ and $A$ correspond to the stiffness of rotor and area of the rotor disc, respectively, 
$W_i$ represents the weight of UAV $i$, 
and
${v}_{z,i}\left[ n \right]=\frac{1}{\delta_t}\left(z_i\left[ n + 1 \right]-z_i\left[ n \right]\right) $ represents the vertical flight speed of UAV $i$. 
It should be noted that UAVs consume no power during vertical descent \cite{FilipponeABook}. Therefore, when $v_{z,i}\left[ n \right] < 0$, we have ${P_{{i}}^{{\mathrm{ver}}}}\left[ n \right]=0$.

In underlay mode, the interference caused by $J$ and ${{D}_{k}}$ must be constrained within the IT threshold of $P$, the following constraint is imposed 
\begin{align}
	&\frac{P_{JP}^{\text{L}}\left[ n \right]{{P}_{J}}\left[ n \right] {{\rho }_{0}}}{d_{JP}^{{{\alpha }_{\text{L}}}}\left[ n \right]} + \frac{P_{JP}^{\text{N}}\left[ n \right]{{P}_{J}}\left[ n \right] {{\rho }_{0}} \mu}{d_{JP}^{{{\alpha }_{\text{N}}}}\left[ n \right]} \nonumber \\
	&\quad \quad \quad \quad \quad \quad  + \sum\limits_{k=1}^{K} {a_k}\left[ n \right]{P_k}\left[ n \right]{\mathbb{E}}\left[ {{{ {{{\hat h}_{kP}}} }}} \right] \le \Gamma, \forall n, \label{IT}
\end{align}
where 
$P_{JP}^{\text{L}}\left[ n \right]=\frac{1}{1+a{{e}^{-b\left( {{\theta }_{JP}}\left[ n \right]-a \right)}}}$, 
${{\theta }_{JP}}\left[ n \right] = \frac{180}{\pi} \arctan\left( \frac{{{z}_{J}}\left[ n \right]}{\left\| {{\mathbf{q}}_{J}}\left[ n \right] - {{\mathbf{w}}_{P}} \right\|} \right)$, 
$P_{JP}^{\text{N}}\left[ n \right] = 1 - P_{JP}^{\text{L}}\left[ n \right]$,
$h_{JP}^{\text{L}}\left[ n \right] = {{\rho }_{0}} d_{JP}^{-{{\alpha }_{\text{L}}}}\left[ n \right]$ and $h_{JP}^{\text{N}}\left[ n \right] = \mu {{\rho }_{0}} d_{JP}^{-{{\alpha }_{\text{N}}}}\left[ n \right]$, 
${{d}_{JP}}\left[ n \right] = \sqrt{{{z}_{J}}{{\left[ n \right]}^{2}} + {{\left\| {{\mathbf{q}}_{J}}\left[ n \right] - {{\mathbf{w}}_{P}} \right\|}^{2}}}$, 
${{\hat h}_{kP}} = \frac{{{\rho }_{0}}}{d_{kP}^{{{\alpha }_{P}}}} {{\xi }_{kP}}$ signifies the channel power gains between ${{D}_{k}}$ and $P$, 
$d_{kP}$ denotes the distance between $D_k$ and $P$, ${{\alpha }_{P}}$ is the path loss exponent, 
${{\xi }_{kP}}$ is a random variable following an exponential distribution with unit mean, 
and 
$\Gamma$ is the IT threshold at $P$.

\section{Problem Formulation}
\label{Problem}

In this study, the objective is to maximize the ASSE by optimizing the 3D trajectory of the UAVs, the transmit power, and the user scheduling. 
Let 
$\mathbf{A}=\left\{a_k\left[ n \right], \forall k, n\right\}$, 
$\mathbf{P}= \left\{P_k\left[ n \right], P_J\left[ n \right], \forall k, n\right\}$, 
$\mathbf{Q}=\left\{\mathbf{q}_U\left[ n \right], \mathbf{q}_J\left[ n \right], \forall n\right\}$, 
$\mathbf{H}=\left\{z_U\left[ n \right], z_J\left[ n \right], \forall n\right\}$, 
and 
$\mathbf{\Theta}=\left\{\theta_{kU}\left[ {n} \right], \theta_{JP}\left[ {n} \right], \forall n,k\right\}$, 
the following optimization problem is formulated
\begin{subequations}
	\begin{align}
		\mathcal{P}_1: & \max _{\mathbf{A},\mathbf{P},\mathbf{\Theta}, \mathbf{Q}, \mathbf{H}} {\frac{1}{N} \sum_{n=1}^{N} \sum_{k=1}^K a_k\left[ n \right] \bar{R}^{k}_{\sec}\left[ n \right]} \label{p01a}\\
		\mathrm {s.t.} 
		& \frac{1}{N} \sum_{n=1}^{ N} \sum_{k=1}^K a_k\left[ n \right] P_k\left[ n \right] \leq P_{K}^{\mathrm{ave}}, 						\label{p01b} \\
		& 0 \leq P_k\left[ n \right] \leq P_{K}^{\max}, \forall k, \label{p01c}\\
		&\frac{1}{N} \sum\limits_{n=1}^N {P_{{i}}^{{\mathrm{hor}}}}\left[ n \right] \leq {P_{{i,\mathrm {ave }}}^{{\mathrm{hor}}}}, \forall i, \label{p01d}\\
		&\frac{1}{N} \sum\limits_{n=1}^N {P_{{i}}^{{\mathrm{ver}}}}\left[ n \right] \leq {P_{{i,\mathrm {ave }}}^{{\mathrm{ver}}}}, \forall n, i,   \label{p01e}\\		
		& 0 \leq \frac{1}{N} \sum\limits_{n=1}^{N} P_J\left[ n \right] \leq {P_J^{\mathrm{ave}}},  									\label{p01f} \\
		& 0 \leq P_J\left[ n \right] \leq P_J^{\max },   \forall n, \label{p01g}\\
		& \theta_{kU}\left[ n \right]=\frac{180}{\pi} \arctan \left(\frac{z_U\left[ n \right]}{\left\|\mathbf{q}_U\left[ n \right]-\mathbf{w}_{k}\right\|}\right),  \forall n, k, 												\label{p01h}\\
		& \theta_{JP}\left[ n \right]=\frac{180}{\pi} \arctan \left(\frac{z_J\left[ n \right]}{\left\|\mathbf{q}_J\left[ n \right]-\mathbf{w}_{P}\right\|}\right),  \forall n, 												\label{p01i}\\		
		& \left|\mathbf{q}_i\left[ {n + 1} \right]-\mathbf{q}_i\left[ n \right]\right| = \delta_t {\mathbf v}_{xy,i}\left[ n \right], \forall n, \label{p01j}\\
		&\left\|\mathbf{v}_{xy,i}\left[ n \right]\right\| \leq V_{xy }, \forall n, \label{p01k} \\
		& \left|z_i\left[ {n + 1} \right]-z_i\left[ n \right]\right| = \delta_t v_{z,i}\left[ n \right], \forall n, \label{p01l}\\
		&\left|{v}_{z,i}\left[ n \right]\right| \leq V_z, \forall n, \label{p01m} \\
		&\mathbf{q}_i\left[ 1 \right] = \mathbf{q}_i^{\mathrm I}, \quad \mathbf{q}_i\left[ N \right] = \mathbf{q}_i^{\mathrm F},  								\label{p01n}\\
		&z_i\left[ 1 \right] = z_i^{\mathrm I}, \quad z_i\left[ N \right] = z_i^{\mathrm F}, \label{p01o} \\
		& H_{\min } \leq z_i\left[ n \right] \leq H_{\max }, \forall n,    \label{p01p}\\
		& \sqrt{\left\|\mathbf{q}_{i_1}\left[ n \right]-\mathbf{q}_{j_1}\left[ n \right]\right\|^2+\left|z_{i_1}\left[ n \right]-z_{j_1}\left[ n \right]\right|^2} \geq D_{\min },    \nonumber\\
		& \forall n, \forall {i_1}, {j_1} \in\{U, J, E\}, {i_1} \neq {j_1}, \label{p01q}\\
		&{{\frac{1}{N} \sum_{n=1}^{N}  a_k\left[ n \right] \bar{R}^{k}_{\sec}\left[ n \right]} \geq R_{\min} ,\forall k,}  \label{p01r} \\
		&(\rm{\ref{diaodu1}}),  (\rm{\ref{diaodu2}}), (\rm{\ref{IT}}), \nonumber
	\end{align}
\end{subequations}
where 
$P_{K}^{\mathrm{ave}}$ and $P_{K}^{\max}$ denote the average and maximum transmit power of GDs, respectively, 
${P_{{i,\mathrm {ave }}}^{{\mathrm{hor}}}}$ and ${P_{{i,\mathrm {ave }}}^{{\mathrm{ver}}}}$ represent the average transmit powers of $i$, respectively, 
$P_{J}^{\text{ave}}$ and $P_{J}^{\max}$ represent the average and maximum jamming power for $J$, respectively, 
$V_{xy }$ and $V_z$ represent the maximum horizontal and vertical velocities, respectively,
$\mathbf{q}_i^{\mathrm I}$ and $\mathbf{q}_i^{\mathrm F}$ represent the initial and the final horizontal positions of UAVs, 
$z_i^{\mathrm I}$ and $z_i^{\mathrm F}$ denote the initial and the final vertical positions of UAVs, 
$H_{\min}$ and $H_{\max}$ denote the minimum and maximum vertical heights of $U$ and $J$, respectively, 
and 
$D_{\min }$ represents the minimum safe distance between UAVs.
In particular, 
($\rm{\ref{p01b}}$) and ($\rm{\ref{p01c}}$) refer to the average and maximum transmit power of $D_k$, respectively,
($\rm{\ref{p01d}}$) and ($\rm{\ref{p01e}}$) establish the limits on propulsion energy consumption for $U$ and $J$, 
($\rm{\ref{p01f}}$) and ($\rm{\ref{p01g}}$) outline the limits on the average and peak jamming power for each slot,
{($\rm{\ref{p01h}}$) specifies the required elevation angle between $D_k$ and $U$, ($\rm{\ref{p01i}}$) specifies the required elevation angle between $J$ and $P$,}
($\rm{\ref{p01j}}$)-($\rm{\ref{p01m}}$) are the velocity constraints, 
($\rm{\ref{p01n}}$) and ($\rm{\ref{p01o}}$) are 
the initial and final positions constraints, respectively, 
($\rm{\ref{p01p}}$) denotes the flight altitude constraint for UAVs,
($\rm{\ref{p01q}}$) sets the safety distance among UAVs ($U$, $J$, $E$), 
and 
($\rm{\ref{p01r}}$) describes the minimum {SSE} constraint for $D_k$.
 
{Several factors complicate the resolution of the optimization problem $\mathcal{P}_1$. Firstly, the operator $[\cdot]^{+}$ introduces non-smoothness into the objective function when it approaches zero. Secondly, the achievable secrecy rate, as outlined in (\ref{securerate1}), is affected by both the 3D trajectory of the UAV and the probability of LoS versus NLoS communication. }
Moreover, the user scheduling constraint depicted in (\ref{diaodu2}) introduces a binary integer constraint that is inherently non-convex. Additionally, the elevation constraints specified in (\ref{p01h}) and ($\rm{\ref{p01i}}$), which arise from the PLoS model, are also nonlinear. As a consequence, $\mathcal{P}_1$ is classified as a nonlinear mixed-integer non-convex optimization problem, which generally poses considerable difficulties in finding a solution.

\section{Proposed Algorithm for Problem $\mathcal{P}_1$}
\label{ProposedAlgorithm1}

To reduce the complexity, the approximate method in \cite{YouC2020TWC} is utilized in this work. 
Specifically, we only consider the receiving {SE} in the LoS scenario, which can significantly reduce the complexity of the formulated problem. 	
Since the {SE} in NLoS scenarios is much lower than that in LoS scenarios, 
the lower bound for (\ref{expectedRkU}) is approximated as \cite{YouC2020TWC, Lei2024TCCN, MengA2022IoT}
\begin{align}
	\bar{R}_{kU}^{{\mathrm {lb}}}\left[ n \right]=P_{k U}^{\mathrm L}\left[ n \right] \log _2\left(1+\frac{P_k\left[ n \right] h_{k U}^{\mathrm L}\left[ n \right]}{\sigma^2}\right). \label{keda1} 
\end{align}
Then, the achievable {SSE} of $D_k$ over the $n$-th slot is approximated by 
$	{\bar{R}_{\sec}}^k\left[ n \right]=\left[\bar{R}_{kU}^{\mathrm {lb}}\left[ n \right]-\bar{R}_{kE}^{{\mathrm{ub}}}\left[ n \right]\right]^{+}$. 
Then, we address the non-smoothness of the objective function. When the term $\bar{R}_{kU}^{\mathrm{lb}}\left[ n \right] - \bar{R}_{kE}^{\mathrm{ub}}\left[ n \right]$ from (\ref{securerate1}) becomes negative during the $n$-th slot, we set the transmit power $P_k\left[ n \right]$ to zero, which leads to $\bar{R}_{\sec}^k\left[ n \right] = 0$. This adjustment ensures that the objective function remains non-negative, allowing us to remove the operator $[\cdot]^{+}$ without affecting the optimal solution of the original problem.

Next, to solve the optimization problem $\mathcal{P}_1$, we utilize the BCD technique\footnote{
	{It is worth noting that to address high-dimensional, non-convex joint optimization problems, the academic has begun to explore data-driven deep reinforcement learning approaches \cite{AAl2023IWCMC}, \cite{AAl2025OJCS}, \cite{LeiH2024TCCNFair}. However, such methods still face challenges in ensuring theoretical convergence, strictly satisfying complex security constraints, and training costs. In contrast, this work aims to construct a rigorous and analytically processable model-driven optimization framework. Our proposed solution, based on BCD and SCA, aims to provide a performance benchmark and theoretical reference with clear convergence guarantees for this scenario, laying the foundation for subsequent research on fusion learning and optimization paradigms. 
		Moreover, it should be noted that the offline optimization in this work is the first step towards these complex real-time systems, and the motivation lies in proving that under such complex coupling constraints (security, CRN, 3D trajectory, energy consumption), the optimization problem is moldable and solvable, and provides a key benchmark for system performance.
	}
}, 
enabling us to decompose the original problem into several manageable subproblems. In this approach, we optimize the variables $\mathbf{A}$, $\mathbf{P}$, $\mathbf{H}$, $\mathbf{Q}$, and $\mathbf{\Theta}$ in each subproblem while keeping the other variables fixed. Additionally, we apply the SCA method to transform the non-convex constraints into convex ones, thereby simplifying the optimization process. This combined strategy significantly enhances our capability to systematically and efficiently identify sub-optimal solutions.

\subsection{Subproblem 1: Optimizing Scheduling Variable }	
In this subsection, we focus on optimizing the user scheduling variable $\mathbf{A}$ while keeping $\mathbf{P}$, $\mathbf{H}$, $\mathbf{Q}$, and $\mathbf{\Theta}$ fixed. We relax the binary constraint $a_k\left[ n \right] \in \{0,1\}$ to a linear variable, allowing $0 \leq a_k\left[ n \right] \leq 1$. Consequently, $\mathcal{P}_1$ is reformulated as
\begin{subequations}
	\begin{align}
		\mathcal{P}_{2.1}: & \max _{\mathbf{A},\eta_k} \sum_{k=1}^K \eta_k \\
		\mathrm { s.t. } & \frac{1}{N} \sum_{n=1}^{N} a_k\left[ n \right] \bar{R}_{\sec}^{k}\left[ n \right] \geq \eta_k,  \forall k,\\
		& \eta_k \geq R_{\min}, \forall k, \\
		& 0\leq a_k\left[ n \right]\leq1, \forall n, k, \label{p11c}\\
		& (\rm{\ref{diaodu1}}),(\rm{\ref{IT}}),(\rm{\ref{p01b}}). \nonumber
	\end{align}
\end{subequations}
{where 
$\eta_k$ is a slack variable. $\mathcal{P}_{2.1}$ is a strictly linear optimization problem, which can be efficiently addressed using the CVX toolbox, as described in \cite{tuyouhua}.}

\subsection{Subproblem 2: Optimizing Transmit Power of $D_k$ and $J$}

{
In this subsection, $\mathbf{P}$  is optimized for given $\mathbf{A}$, $\mathbf{Q}$, $\mathbf{H}$, and $\mathbf{\Theta}$. $\mathcal{P}_1$ is reformulated as
\begin{subequations}
	\begin{align}
		\mathcal{P}_{2.2\rm{a}}: & \max _{\mathbf{P},\eta_k} \sum_{k=1}^K \eta_k \\
		\mathrm { s.t. } 
		& \frac{1}{N} \sum_{n=1}^{N} a_k\left[ n \right] \bar{R}_{\sec}^{k}\left[ n \right] \geq \eta_k,  \forall k, \label{sp2} \\ 
		& \eta_k \geq R_{\min}, \forall k, \\
		& (\rm{\ref{IT}}),(\rm{\ref{p01b}}), (\rm{\ref{p01c}}), (\rm{\ref{p01f}}),(\rm{\ref{p01g}}). \nonumber
	\end{align}
\end{subequations}}
To discuss the convex of $\bar{R}_{\sec}^{k}\left[ n \right]$, we rewrite it as
\begin{align}
	\bar{R}_{\sec}^{k}\left[ n \right]& =  P_{k U}^{{\mathrm L}}\left[ n \right] \log _2\left(1+\frac{P_k\left[ n \right] h_{k U}^{{\mathrm L}}\left[ n \right]}{\sigma^2}\right)\nonumber \\
	& - P_{k E}^{{\mathrm L}}\left[ n \right] \log _2\left(P_k\left[ n \right] h_{k E}^{{\mathrm L}}\left[ n \right] + P_J\left[ n \right] h_{J E}\left[ n \right]+\sigma^2\right) \nonumber\\
	& + P_{k E}^{{\mathrm L}}\left[ n \right] \log _2\left(P_J\left[ n \right] h_{J E}\left[ n \right]+\sigma^2\right) \nonumber \\ 
	& - P_{k E}^{{\mathrm N}}\left[ n \right] \log _2\left(P_k\left[ n \right] h_{k E}^{{\mathrm N}}\left[ n \right] + P_J\left[ n \right] h_{J E}\left[ n \right]+\sigma^2\right) \nonumber\\
	& + P_{k E}^{{\mathrm N}}\left[ n \right] \log _2\left(P_J\left[ n \right] h_{J E}\left[ n \right]+\sigma^2\right). 
	\label{zhankai1}
\end{align}
It is easy to note that 
the second and fourth terms in (\ref{zhankai1}) are convex functions 
with respect to $\mathbf{P}$, which make (\ref{sp2}) a non-convex constraint. 
To tackle this non-convex constraint, we utilize a first-order Taylor expansion to approximate $\bar{R}_{\sec}^{k}\left[ n \right]$ as $\bar{R}_{\sec,1}^{k, {\mathrm {lb}}}\left[ n \right]$, which is found as 
(\ref{zhankai2}), shown at the top of the next page,
\begin{figure*}[ht]
	\begin{align}
		\bar{R}_{\sec, 1}^{k, {\mathrm {lb}}}\left[ n \right] &=  P_{k U}^{{\mathrm L}}\left[ n \right] \log _2\left(1+\frac{P_k\left[ n \right] h_{k U}^{{\mathrm L}}\left[ n \right]}{\sigma^2}\right)  + \log _2\left(P_J\left[ n \right] h_{J E}\left[ n \right]+\sigma^2\right) \nonumber \\
		& - P_{k E}^{{\mathrm L}}\left[ n \right] \log _2\left(P_k^{\left( m \right)}\left[ n \right] h_{k E}^{{\mathrm L}}\left[ n \right] + P_J^{\left( m \right)}\left[ n \right] h_{J E}\left[ n \right]+\sigma^2\right) \nonumber \\
		& - P_{k E}^{{\mathrm N}}\left[ n \right] \log _2\left(P_k^{\left( m \right)}\left[ n \right] h_{k E}^{{\mathrm N}}\left[ n \right] + P_J^{\left( m \right)}\left[ n \right] h_{J E}\left[ n \right]+\sigma^2\right) \nonumber \\
		&  - P_{k E}^{{\mathrm L}}\left[ n \right] \frac{h_{k E}^{{\mathrm L}}\left[ n \right] (P_k\left[ n \right] - P_k^{\left( m \right)}\left[ n \right]) + h_{J E}\left[ n \right] (P_J\left[ n \right] - P_J^{\left( m \right)}\left[ n \right])}{\ln(2) \left(P_k^{\left( m \right)}\left[ n \right] h_{k E}^{{\mathrm L}}\left[ n \right] + P_J^{\left( m \right)}\left[ n \right] h_{J E}\left[ n \right]+\sigma^2\right)}  \nonumber \\
		&  - P_{k E}^{{\mathrm N}}\left[ n \right] \frac{h_{k E}^{{\mathrm N}}\left[ n \right] (P_k\left[ n \right] - P_k^{\left( m \right)}\left[ n \right]) + h_{J E}\left[ n \right] (P_J\left[ n \right] - P_J^{\left( m \right)}\left[ n \right])}{\ln(2) \left(P_k^{\left( m \right)}\left[ n \right] h_{k E}^{{\mathrm N}}\left[ n \right] + P_J^{\left( m \right)}\left[ n \right] h_{J E}\left[ n \right]+\sigma^2\right)}  \label{zhankai2}
	\end{align}
	\hrulefill
\end{figure*}
where $\{P_k^{\left( m \right)}\left[ n \right], P_J^{\left( m \right)}\left[ n \right], \forall n\}$ denotes a feasible point chosen during the $m$-th iteration. 
By substituting $\bar{R}_{\sec}^{k}\left[ n \right]$ with $\bar{R}_{\sec,1}^{k, {\mathrm {lb}}}\left[ n \right]$, $\mathcal{P}_{2.2 \rm{a}}$ is reformulated into a convex optimization problem, which is expressed as 
\begin{subequations}
	\begin{align}
		\mathcal{P}_{2.2\rm{b}}: & \max _{\mathbf{P},\eta_k}\sum_{k=1}^K  \eta_k \\
		\mathrm { s.t. } 
		& \frac{1}{N} \sum_{n=1}^{N} a_k\left[ n \right] \bar{R}_{\sec,1}^{k,{\mathrm {lb}}}\left[ n \right] \geq \eta_k, \forall k, \\ 
		& \eta_k \geq R_{\min}, \forall k, \\
		& (\rm{\ref{IT}}),(\rm{\ref{p01b}}), (\rm{\ref{p01c}}), (\rm{\ref{p01f}}),(\rm{\ref{p01g}}). \nonumber
	\end{align}
\end{subequations}

By employing CVX, we can effectively address this optimization problem. This method not only streamlines the solving process but also enhances the chances of identifying the optimal solution.

\subsection{Subproblem 3: Optimizing Horizontal Trajectory of $U$ and $J$ }
In this subsection, we optimize the horizontal trajectory of $ U $ and $ J $ with given $\{\mathbf{A}, \mathbf{P}, \mathbf{H}\}$. $\mathcal{P}_1$ is reformulated as 
\begin{subequations}
	\begin{align}
		\mathcal{P}_{2.3 \rm{a}}: & \max _{\mathbf{Q}, \mathbf{\Theta},\eta_k} \sum_{k=1}^K \eta_k \\
		\mathrm { s.t. } & \frac{1}{N} \sum_{n=1}^{ N} a_k\left[ n \right] \bar{R}_{\sec}^{k}\left[ n \right] \geq \eta_k, \forall k, \label{P23b}	\\
			& \eta_k \geq R_{\min}, \forall k, \\
		&(\rm{\ref{IT}}), (\rm{\ref{p01d}}), (\rm{\ref{p01h}})-(\rm{\ref{p01k}}), (\rm{\ref{p01n}}), (\rm{\ref{p01q}}).		\nonumber
	\end{align}
\end{subequations}
It should be noted that 
$\bar{R}_{\sec}^{k}\left[ n \right]$ in constraint (\ref{P23b}) is affected not only by the variable $\mathbf{Q}$ but also by $P_{kU}^{{\mathrm L}}\left[ n \right]$. 
Moreover, $\mathcal{P}_{2.3 \rm{a}}$ is non-convex due to the inclusion of constraints (\ref{IT}), (\ref{p01d}),  (\ref{p01h}), (\ref{p01i}), and (\ref{p01q}). 
In conclusion, the non-convex nature stems from both the specific constraints that fail to meet convexity requirements and the complex interactions among the variables involved.

1) First, since the left-hand side (LHS) of $\bar{R}_{\sec}^k\left[ n \right]$ in (\ref{P23b}) is non-concave, the optimization problem becomes non-convex, presenting significant challenges for resolution. 
To simplify the derivation, we introduce auxiliary variables
$Y\left[ n \right] = \frac{\rho_0}{\sigma^2 \left( \left(\left\|\mathbf{q}_J\left[ n \right]-\hat{\mathbf{q}}_E\left[ n \right]\right\| + r_E\right)^2 + \left| z_J\left[ n \right] - z_E\left[ n \right] \right|^2 \right)}$, 
$t_{kU}\left[ n \right]=  \left( \|\mathbf{q}_U\left[ n \right] - \mathbf{w}_k\|^2 + z_U\left[ n \right]^2 \right)^{\frac{\alpha_{{\mathrm L}}}{2}}$, 
and
$x_k\left[ n \right] = 1+a e^{-b\left(\theta_{kU}\left[ n \right]-a\right)}$. 
Then, $\bar{R}_{\sec}^k\left[ n \right]$ in (\ref{P23b}) is expressed as
\begin{align}
	\bar{R}_{\sec,2}^k\left[ n \right]& = \frac{1}{x_k\left[ n \right]} \log_2\left(1 + \frac{P_k\left[ n \right]  \gamma}{t_{kU}\left[ n \right]}\right) \nonumber \\
	& - P_{kE}^{{\mathrm L}}\left[ n \right] \log _2\left(1+\frac{b_{kL}\left[ n \right]}{P_{J}\left[ n \right] Y\left[ n \right]+1}\right) \nonumber \\ 
	& - P_{kE}^{{\mathrm N}}\left[ n \right] \log _2\left(1+\frac{b_{kN}\left[ n \right]}{P_{J}\left[ n \right] Y\left[ n \right]+1}\right),
	\label{secure2}
\end{align}
where 
$\gamma = \frac{\rho_0}{\sigma^2}$, $b_{kL}\left[ n \right]= \frac{P_k\left[ n \right]h_{kE}^{{\mathrm L},{ub}}\left[ n \right]}{\sigma^2}$, and $b_{kN}\left[ n \right]= \frac{P_k\left[ n \right]h_{kE}^{{\mathrm N},{ub}}\left[ n \right]}{\sigma^2}$. 
For $Y\left[ n \right]$, the additional constraint in (\ref{yy}), shown at the top of the next page, must be satisfied. 
\setcounter{equation}{26} 
\begin{figure*}[ht]
	\begin{align}
		\frac{1}{Y\left[ n \right]} \geq \frac{\sigma^2\left(\left(\left\|\mathbf{q}_J\left[ n \right]-\hat{\mathbf{q}}_E\left[ n \right]\right\| + r_E\right)^2+ | z_J\left[ n \right] - z_E\left[ n \right] |^2\right)}{\rho_0}, \forall n
		\label{yy}
	\end{align}
	\hrulefill
\end{figure*}
Then after performing SCA convex transformation on the LHS of (\ref {yy}), we obtain 
\begin{align}
	\frac{2 {Y}^{\left( m \right)}\left[ n \right]-Y\left[ n \right]}{({Y^{\left( m \right)}\left[ n \right]})^2} &\geq \left(\left\|\mathbf{q}_J\left[ n \right]-\hat{\mathbf{q}}_E\left[ n \right]\right\| + r_E\right)^2
	\nonumber \\ 
	&+ \left| z_J\left[ n \right] - z_E\left[ n \right] \right|^2, 
	\label{yySCA}
\end{align}
where 
$Y^{\left( m \right)}\left[ n \right]$ denotes $Y\left[ n \right]$ at the $m$-th iteration 
and 
$Y^{\left( m \right)}\left[ n \right] = \frac{\rho_0}{\sigma^2 \left( \left(\left\|\mathbf{q}_J^{\left( m \right)}\left[ n \right]-\hat{\mathbf{q}}_E\left[ n \right]\right\| + r_E\right)^2
	+ \left| z_J^{\left( m \right)}\left[ n \right] - z_E\left[ n \right] \right|^2 \right)}$.
	Finally, according to Appendix A in \cite{Lei2024TCCN}, for $A \geq 0$, the function $f(x, y) = \frac{1}{x} \log_2\left(1 + \frac{A}{y}\right)$ is jointly convex with respect to $x > 0$ and $y > 0$. Consequently, $\bar{R}_{\sec,2}^k\left[ n \right]$ in (\ref{secure2}) is approximated by the following lower bound
	\begin{align}
		\bar{R}_{\sec,2}^{k, {\mathrm {lb}}}\left[ n \right] & = \frac{1}{x_k^{\left( m \right)}\left[ n \right]}  \log_2\left(1 + \frac{P_k\left[ n \right] \gamma}{t_{kU}^{\left( m \right)}\left[ n \right]}\right) \nonumber \\
		& - \frac{P_k\left[ n \right] \gamma (t_{kU}\left[ n \right] - t_{kU}^{\left( m \right)}\left[ n \right])}{\ln(2) (x_k^{\left( m \right)}\left[ n \right]) (t_{kU}^{\left( m \right)}\left[ n \right])^2 \left(1 + \frac{P_k\left[ n \right] \gamma}{t_{kU}^{\left( m \right)}\left[ n \right]}\right)} \nonumber \\
		& - \frac{\log_2\left(1 + \frac{P_k\left[ n \right] \gamma}{t_{kU}\left[ n \right]} \right)(x_k\left[ n \right] - x_k^{\left( m \right)}\left[ n \right])}{\ln(2) (x_k^{\left( m \right)}\left[ n \right])^2} \nonumber \\
		& - P_{kE}^{{\mathrm L}}\left[ n \right] \log _2\left(1+\frac{b_{kL}\left[ n \right]}{p_{J}\left[ n \right] Y\left[ n \right]+1}\right) \nonumber \\
		& - P_{kE}^{{\mathrm N}}\left[ n \right] \log _2\left(1+\frac{b_{kN}\left[ n \right]}{p_{J}\left[ n \right] Y\left[ n \right]+1}\right),
		\label{secure2lb}
	\end{align}
	where 
	$x_k^{\left( m \right)}\left[ n \right]$ and $t_{kU}^{\left( m \right)}\left[ n \right]$ signify $x_k\left[ n \right]$ and $t_{kU}\left[ n \right]$ at the $m$-th iteration, 
	$x_k^{\left( m \right)}\left[ n \right] = 1+a e^{-b\left(\theta_{kU}^{\left( m \right)}\left[ n \right]-a\right)}$, $\theta_{k U}^{\left( m \right)}\left[ n \right]= \frac{180}{\pi} \arctan \left(\frac{z_U\left[ n \right]}{\left\|\mathbf{q}_U^{\left( m \right)}\left[ n \right]-\mathbf{w}_k\right\|}\right)$, and 
	$t_{kU}^{\left( m \right)}\left[ n \right]=  \left( \|\mathbf{q}_U^{\left( m \right)}\left[ n \right] - \mathbf{w}_k\|^2 + z_U\left[ n \right]^2 \right)^{\frac{\alpha_{{\mathrm L}}}{2}}$.

\setcounter{equation}{32}
\begin{figure*}[ht]
	\begin{subequations}
		\begin{align}
			\frac{1}{N}\sum\limits_{n = 1}^N {\left[ {{P_0}\left( {1 + \frac{{3{{\left\| {{\mathbf{v}}_{xy, i}\left[ n \right]} \right\|}^2}}}{{U_{{\mathrm{tip}}}^2}}} \right) + \frac{1}{2}{d_0}\rho sA{{\left\| {{\mathbf{v}}_{xy, i}\left[ n \right]} \right\|}^3} + {P_1}{\lambda _i}\left[ n \right]} \right]}  \le {P_{{i,\mathrm {ave }}}^{{\mathrm{hor}}}}, \forall n, i \in\{U, J\} 	\label{shuipingslack} \\
			\frac{1}{\lambda_{{i}}\left[ n \right]^2} \leq \lambda_{{i}}\left[ n \right]^2+\frac{\left\|{\mathbf{v}}_{xy, i}\left[ n \right]\right\|^2}{v_0^2},  \forall n, i \in\{U, J\} 	\label{lambda}
		\end{align}
	\end{subequations}
	\hrulefill
\end{figure*}

2) 
For (\ref{IT}) and (\ref{p01i}), similar to \cite{Lei2024TCCN}, by introducing slack variables 
${{\mathbf{p}}_{h}}=\left\{ {{p}_{h}}\left[ n \right],\forall n \right\}$ 
and 
${{\mathbf{t}}_{P,h}}=\left\{ {{t}_{P,h}}\left[ n \right],\forall n \right\}$, where $h\in \{\text{L},\mathbf{N}\}$, (\ref{IT}) is reformulated as  \setcounter{equation}{29}
\begin{align}
	\frac{{{P}_{J}}\left[ n \right]{{\rho }_{0}}}{{{p}_{\text{L}}}\left[ n \right]{{t}_{P,\text{L}}}\left[ n \right]}+\frac{{{P}_{J}}\left[ n \right]{{\rho }_{0}}\mu }{{{p}_{\text{N}}}\left[ n \right]{{t}_{P,\text{N}}}\left[ n \right]} + \sum\limits_{k=1}^{K} {\frac{{{a}_{k}}\left[ n \right]{{P}_{k}}\left[ n \right] {{\rho }_{0}}}{d_{kP}^{{{\alpha }_{P}}}}} \le \Gamma,
	\label{ITnew1}
\end{align}
with 
\begin{subequations}
	\begin{align}
		p_\mathrm{L}\left[ n \right] &\leq 1+a e^{a b} e^{-b \theta_{JP}\left[ n \right]} \label{A}\\
		\theta_{JP}\left[ n \right] &\geq \frac{180}{\pi} \arctan \left(\frac{z_J\left[ n \right]}{\left\|\mathbf{q}_J\left[ n \right]-\mathbf{w}_P\right\|}\right) \label{B}\\
		p_\mathrm{N}\left[ n \right] &\leq 1+\frac{1}a e^{-a b} e^{b \phi_{JP}\left[ n \right]} \label{C}\\
		\phi_{JP}\left[ n \right] &\leq \frac{180}{\pi} \arctan \left(\frac{z_J\left[ n \right]}{\left\|\mathbf{q}_J\left[ n \right]-\mathbf{w}_P\right\|}\right) \label{D} \\
		t_{P,h}\left[ n \right] &\leq\left(\left\|\mathbf{q}_J\left[ n \right]-\mathbf{w}_{P}\right\|^2+z_J^2\left[ n \right]\right)^{\alpha_h / 2}. \label{E}
	\end{align}
\end{subequations}
By utilizing SCA, non-convex constraints, (\ref{A}), (\ref{C}), (\ref{D}), and (\ref{E}) are transformed as 
\begin{subequations}
	\begin{align}
		p_\mathrm{L}\left[ n \right] &\leq 1+a e^{ab}\kappa_1^{{\left(m  \right)}}\left[ n \right], \label{Aconvex}\\
		p_\mathrm{N}\left[ n \right] &\leq 1+\frac{1}{a} e^{-ab}\kappa_2^{{\left( m \right)}}\left[ n \right],\label{Cconvex} \\
		\phi_{JP}\left[ n \right] &\leq \frac{180}{\pi}\kappa_3^{{\left( m  \right)}}\left[ n \right],\label{Dconvex} \\
		t_{P, h}\left[ n \right] &\leq \kappa_{4,h}^{{\left( m  \right)}}\left[ n \right], \label{Econvex}
	\end{align}
\end{subequations}
where 
$\theta _{JP}^{\left( m \right)}\left[ n \right]$, $\phi _{JP}^{\left( m \right)}\left[ n \right]$, $d_{JP}^{\left( m \right)}\left[ n \right]$, and $\mathbf{q}_{J}^{\left( m \right)}\left[ n \right]$ respectively represent the values of ${\theta _{JP}}\left[ n \right]$, ${\phi _{JP}}\left[ n \right]$, ${d_{JP}}\left[ n \right]$, and ${\mathbf{q}_{J}}\left[ n \right]$ at the $m$-th iteration,
$\kappa_1^{\left( m \right)}\left[ n \right]={{e}^{-b\theta _{JP}^{\left( m \right)}\left[ n \right]}}-b{{e}^{-b\theta _{JP}^{\left( m \right)}\left[ n \right]}}\left( {{\theta }_{JP}}\left[ n \right]-\theta _{JP}^{\left( m \right)}\left[ n \right] \right)$, 
$\kappa_2^{\left( m \right)}\left[ n \right] ={{e}^{b\phi _{JP}^{\left( m \right)}\left[ n \right]}}+b{{e}^{b\phi _{JP}^{\left( m \right)}\left[ n \right]}}\left( {{\phi }_{JP}}\left[ n \right]-\phi _{JP}^{\left( m \right)}\left[ n \right] \right)$, 
$\kappa_3^{\left( m \right)}\left[ n \right] = \arctan \left( \frac{{{z}_{J}}\left[ n \right]}{\left\| \mathbf{q}_{J}^{\left( m \right)}\left[ n \right]-{{\mathbf{w}}_{P}} \right\|} \right) -\frac{{{z}_{J}}\left[ n \right]\left( \left\| {{\mathbf{q}}_{J}}\left[ n \right]-{{\mathbf{w}}_{P}} \right\|-\left\| \mathbf{q}_{J}^{\left( m \right)}\left[ n \right]-{{\mathbf{w}}_{P}} \right\| \right)}{{{\left\| \mathbf{q}_{J}^{\left( m \right)}\left[ n \right]-{{\mathbf{w}}_{P}} \right\|}^{2}}+z_{J}^{2}\left[ n \right]}$, 
and 
$\kappa_{4,h}^{\left( m \right)}\left[ n \right] = {{\left( d_{JP}^{\left( m \right)}\left[ n \right] \right)}^{{{\alpha }_{h}}}}+{{\alpha }_{h}}{{\left( d_{JP}^{\left( m \right)}\left[ n \right] \right)}^{{{\alpha }_{h}}-2}} + {{\left( \mathbf{q}_{J}^{\left( m \right)}\left[ n \right]-{{\mathbf{w}}_{P}} \right)}^{T}}\left( {{\mathbf{q}}_{J}}\left[ n \right]-\mathbf{q}_{J}^{\left( m \right)}\left[ n \right] \right)$. 

3) 
To tackle (\ref{p01d}), we introduce a relaxation variable $\{\lambda_{i}\left[ n \right] \geq 0, i \in \{U, J\}\}$, which allows us to reformulate (\ref{p01d}) into (\ref{shuipingslack}) and (\ref{lambda}), as shown at the top of this page. It is crucial to emphasize that the right-hand side (RHS) of (\ref{lambda}) is a convex function with respect to both $\lambda_{i}\left[ n \right]$ and $\|{\mathbf{v}}_{xy, i}\left[ n \right]\|$, meaning it does not constitute a convex constraint. By employing the SCA method and replacing the RHS with its convex lower bound, we express (\ref{lambda}) as \setcounter{equation}{33}
\begin{align}\label{lambdasca}
	\frac{1}{\lambda_{{i}}\left[ n \right]^2} &\leq {\lambda}^{{\left( m  \right)}}_{{i}}\left[ n \right]^2 + 2 {\lambda}^{{\left( m  \right)}}_{{i}}\left[ n \right]\left(\lambda_{{i}}\left[ n \right]-{\lambda}^{{\left( m  \right)}}_{{i}}\left[ n \right]\right)  \nonumber\\
	& + \frac{2}{v_0^2}\left({\mathbf{v}}^{{\left( m  \right)}}_{xy, i}\left[ n \right]\right)^T\left({\mathbf{v}}_{xy, i}\left[ n \right]-{\mathbf{v}}^{{\left( m  \right)}}_{xy, i}\left[ n \right]\right) \nonumber\\
	& +\frac{\left\|{\mathbf{v}}^{{\left( m  \right)}}_{xy, i}\left[ n \right]\right\|^2}{v_0^2},
\end{align}
where
${\mathbf{v}}^{{\left( m  \right)}}_{xy, i}\left[ n \right]$ represents the horizontal velocity obtained in the $m$-th iteration.

4) 
Then, we address nonlinear constraint (\ref{p01h}). 
Similar to \cite{YouC2020TWC}, we relax it into 
\begin{subequations}
	\begin{align}
	\theta_{kU}\left[ n \right]&\leq\frac{180}{\pi} \arctan \left(\frac{z_U\left[ n \right]}{\left\|\mathbf{q}_U\left[ n \right]-\mathbf{w}_{k}\right\|}\right),  \forall n, k.
	\label{jiaodusongchi1} 
	\end{align}
\end{subequations}
It is noteworthy that the function $\arctan(1/x)$ is convex. As a result, the right-hand side (RHS) of (\ref{jiaodusongchi1}) demonstrate convexity concerning the expression $\left\|\mathbf{q}_U\left[ n \right]-\mathbf{w}_k\right\|$. 
By utilizing the SCA technique, we rewrite (\ref{jiaodusongchi1}) as
\begin{align}
	\theta_{kU}\left[ n \right] &\leq \frac{180}{\pi}\kappa_5^{\left( m \right)}\left[ n \right], \label{F1}
\end{align}
where 
$\kappa_5^{\left( m \right)}\left[n\right] = \arctan\left(\frac{{z_U\left[n\right]}}{{\left\| {\mathbf{q}_U^{\left( m \right)}\left[n\right] - \mathbf{w}_{k}} \right\|}}\right) - \frac{{z_U\left[n\right]\left(\left\| {\mathbf{q}_U\left[n\right] - \mathbf{w}_{k}} \right\| - \left\| {\mathbf{q}_U^{\left( m \right)}\left[n\right] - \mathbf{w}_{k}} \right\| \right)}}{{\left\| {\mathbf{q}_U^{\left( m \right)}\left[n\right] - \mathbf{w}_{k}} \right\|^2 + z_U^2\left[n\right]}}$ and 
$\mathbf{q}_U^{\left( m \right)}\left[ n \right]$ denotes $\mathbf{q}_U\left[ n \right]$ at the $m$-th iteration.

5) 
Next, to handle (\ref {p01q}), we firstly rewrite it as
\begin{align}
	{\left\|\mathbf{q}_{i_1}\left[ n \right]-\mathbf{q}_{j_1}\left[ n \right]\right\|^2+\left|z_{i_1}\left[ n \right]-z_{j_1}\left[ n \right]\right|^2} &\geq D_{\min }^2,  \nonumber\\
	\forall n, \forall {i_1}, {j_1} \in\{U, J, E\}, &{i_1} \neq {j_1}. \label{juli2}
\end{align}
The SCA method is utilized to obtain the lower bound of the norm-squared function in (\ref{juli2}) as 
(\ref{Da})-(\ref{Dc}), shown at the top of the next page, 
\begin{figure*}[ht]
	\begin{subequations}
	\begin{align}
		\| \mathbf{q}_J^{\left( m \right)}\left[ n \right] - \mathbf{q}_E\left[ n \right] \|^2 + 2 \left( \mathbf{q}_J^{\left( m \right)}\left[ n \right] - \mathbf{q}_E\left[ n \right] \right)^T \left( \mathbf{q}_{J}\left[ n \right] - \mathbf{q}_J^{\left( m \right)}\left[ n \right] \right)  + | z_J\left[ n \right] - z_E\left[ n \right] |^2 &\geq D^2_{\min}  				\label{Da}\\
		\| \mathbf{q}_J^{\left( m \right)}\left[ n \right] - \mathbf{q}_U^{\left( m \right)}\left[ n \right] \|^2 + 2 \left( \mathbf{q}_J^{\left( m \right)}\left[ n \right] - \mathbf{q}_U\left[ n \right] \right)^T \left( \mathbf{q}_{J}\left[ n \right] - \mathbf{q}_J^{\left( m \right)}\left[ n \right] \right)	\nonumber \\
		- 2 \left( \mathbf{q}_J\left[ n \right] - \mathbf{q}_U^{\left( m \right)}\left[ n \right] \right)^T \left( \mathbf{q}_{U}\left[ n \right] - \mathbf{q}_U^{\left( m \right)}\left[ n \right] \right) + | z_J\left[ n \right] - z_U\left[ n \right] |^2 &\geq D_{\min}^2 								\label{Db}\\
		\| \mathbf{q}_U^{\left( m \right)}\left[ n \right] - \mathbf{q}_E\left[ n \right] \|^2 + 2 \left( \mathbf{q}_U^{\left( m \right)}\left[ n \right] - \mathbf{q}_E\left[ n \right] \right)^T \left( \mathbf{q}_{U}\left[ n \right] - \mathbf{q}_U^{\left( m \right)}\left[ n \right] \right)   + | z_U\left[ n \right] - z_E\left[ n \right] |^2 &\geq D_{\min}^2 			\label{Dc}
	\end{align}
	\end{subequations}
	\hrulefill
\end{figure*}
where $\mathbf{q}_J^{\left( m \right)}\left[ n \right]$ and $z_J^{\left( m \right)}\left[ n \right]$ represent $\mathbf{q}_J\left[ n \right]$ and $z_J\left[ n \right]$ at the $m$-th iteration. 

Finally, $\mathcal{P}_{2.3 \rm{a}}$ is transformed into 
\begin{subequations}
	\begin{align}
		\mathcal{P}_{2.3{\rm b}}: 
		&\max_{\mathbf{Q}, \mathbf{\Theta}, {\bf{Y}}, {\bm{\lambda }}, {\Phi _{JP}}\atop {{\bf{t}}_{P,h}},{{\bf{p}}_h},{\eta _k}} \sum_{k=1}^K \eta_k \\
		\mathrm { s.t. } 
		&\frac{1}{N} \sum_{n=1}^{N} a_k\left[ n \right] \bar{R}_{\sec,2}^{k, {\mathrm {lb}}}\left[ n \right] \geq \eta_k, \label{sp4} \forall k, \\
		& \eta_k \geq R_{\min}, \forall k, \\
		&(\rm{\ref{p01j}}), (\rm{\ref{p01k}}), (\rm{\ref{p01n}}), (\rm{\ref{yySCA}}), (\rm{\ref{ITnew1}}), (\rm{\ref{B}}), \nonumber\\ &(\rm{\ref{Aconvex}})-(\rm{\ref{Econvex}}), (\rm{\ref{shuipingslack}}), (\rm{\ref{lambdasca}}), (\rm{\ref{F1}}), (\rm{\ref{Da}})-(\rm{\ref{Dc}}), \nonumber 	
	\end{align}
\end{subequations}
{where 
${\bf{Y}} = \left\{ {Y\left[ n \right],\forall n} \right\}$, 
${\bm{\lambda }} = \left\{ {{\lambda _U}\left[ n \right], {\lambda _J}\left[ n \right],\forall n} \right\}$, 
${\Phi _{JP}} = \left\{ {{\phi _{JP}}\left[ n \right],\forall n} \right\}$, 
${{\bf{t}}_{P,h}} = \left\{ {{t_{P,h}}\left[ n \right],\forall n} \right\}$, 
and 
${{\bf{p}}_h} = \left\{ {{p_{\rm{L}}}\left[ n \right],{p_{\rm{N}}}\left[ n \right],\forall n} \right\}$.}
$\mathcal{P}_{2.3 \rm{b}}$ is a convex optimization issue that can be effectively solved using well-known tools such as CVX.

\subsection{Subproblem 4: Optimizing Vertical Trajectory of $U$ and $J$}

In this subsection, we optimize the vertical trajectory of UAVs with given $\{\mathbf{A}, \mathbf{P}, \mathbf{Q}\}$. 
$\mathcal{P}_1$ is reformulated as 
\begin{subequations}
	\begin{align}
		\mathcal{P}_{2.4 \rm{a}}: & \max _{\mathbf{H},\mathbf{\Theta},\eta_k} \sum_{k=1}^K \eta_k \\
		\mathrm { s.t. } 
		& \frac{1}{N} \sum_{n=1}^{ N} a_k\left[ n \right] \bar{R}_{\sec}^{k}\left[ n \right] \geq \eta_k, \label{sp41b} \forall k, \\
		& \eta_k \geq R_{\min}, \forall k, \\
		&(\rm{\ref{IT}}), (\rm{\ref{p01e}}), (\rm{\ref{p01h}}), (\rm{\ref{p01i}}), (\rm{\ref{p01l}}), (\rm{\ref{p01m}}), \nonumber \\
		&(\rm{\ref{p01o}})-(\rm{\ref{p01q}}).  \nonumber
	\end{align}
\end{subequations}

$\mathcal{P}_{2.4 \rm{a}}$ is non-convex because of the constraints (\ref{IT}), (\ref{p01h}), (\ref{p01i}) and (\ref{p01q}). Furthermore, $\bar{R}_{\sec}^{k}\left[ n \right]$ in (\ref{sp41b}) is affected by both $\mathbf{H}$ and $P_{kU}^{{\mathrm L}}\left[ n \right]$.

First, similar to the method in solving the previous subproblem, (\ref{IT}) is reformulated as (\ref{ITnew1}) with additional constraints (\ref{A})-(\ref{E}). 
With respect to $\mathbf{H}$, (\ref{B}) and (\ref{E}) are non-convex. By using SCA, (\ref{B}) and (\ref{E}) are approximated as
\begin{subequations}
	\begin{align}
		{{t}_{P,h}}\left[ n \right]\le \kappa_{6,h}^{(m)}\left[ n \right], \label{F7} \\
		{{\theta }_{JP}}\left[ n \right]\ge \frac{180}{\pi }\kappa_7^{(m)}\left[ n \right], \label{F8}
	\end{align}
\end{subequations}
where 
$z_{J}^{(m)}\left[ n \right]$ and $d_{JP}^{(m)}\left[ n \right]$ represent the values of ${z}_{J}\left[ n \right]$ and $d_{JP}\left[ n \right]$ at the $m$-th iteration, respectively, 
$\kappa_{6,h}^{(m)}\left[ n \right]={{\left( d_{JP}^{(m)}\left[ n \right] \right)}^{{{\alpha }_{h}}}}+{{\alpha }_{h}}{{\left( d_{JP}^{(m)}\left[ n \right] \right)}^{{{\alpha }_{h}}-2}}{{\left( z_{J}^{(m)}\left[ n \right] \right)}^{T}}\left( {{z}_{J}}\left[ n \right]-z_{J}^{(m)}\left[ n \right] \right)$, 
and 
$\kappa_7^{(m)}\left[ n \right]=\arctan \left( \frac{z_{J}^{(m)}\left[ n \right]}{\left\| {{\mathbf{q}}_{J}}\left[ n \right]-{{\mathbf{w}}_{P}} \right\|} \right) +\frac{\left\| {{\mathbf{q}}_{J}}\left[ n \right]-{{\mathbf{w}}_{P}} \right\|\left( {{z}_{J}}\left[ n \right]-z_{J}^{(m)}\left[ n \right] \right)}{{{\left\| {{\mathbf{q}}_{J}}\left[ n \right]-{{\mathbf{w}}_{P}} \right\|}^{2}}+z_{J}^{(m)}{{\left[ n \right]}^{2}}}$.

Then, we address nonlinear constraint (\ref{p01h}) by relaxing it into the following forms
	\begin{align}
		\theta_{kU}\left[ n \right]\leq\frac{180}{\pi} \arctan \left(\frac{z_U\left[ n \right]}{\left\|\mathbf{q}_U\left[ n \right]-\mathbf{w}_{k}\right\|}\right),  \forall n, k, \label{jiaodusongchi} 
	\end{align}
It is important to note that $\arctan(x)$ is a concave function. However, CVX does not include $ \arctan(x) $, so we utilize SCA to convert (\ref{p01h}) as
\begin{align}
	\theta_{kU}\left[ n \right] \leq \frac{180}{\pi}\kappa_8^{\left( m \right)}\left[ n \right], \label{F6} 
\end{align}
where 
$\kappa_8^{\left( m \right)}\left[n\right] = \arctan\left(\frac{z_U^{\left( m \right)}\left[n\right]}{\left\| \mathbf{q}_U\left[n\right] - \mathbf{w}_{k} \right\|}\right) - \frac{\left(\left\| \mathbf{q}_U\left[n\right] - \mathbf{w}_{k} \right\| \right) \left(z_U\left[n\right] - z_U^{\left( m \right)}\left[n\right]\right )}{\left\| \mathbf{q}_U\left[n\right] - \mathbf{w}_{k} \right\|^2 + \left(z_U^{\left( m \right)}\left[n\right]\right)^2}$, and 
$z_U^{\left( m \right)}\left[ n \right]$ denotes $z_U\left[ n \right]$ at the $m$-th iteration.

Next, we rewrite (\ref {p01q}) as
\begin{subequations}
	\begin{align}
		& \| \mathbf{q}_J\left[ n \right] - \mathbf{q}_E\left[ n \right] \|^2 + | z_J^{\left( m \right)}\left[ n \right] - z_E\left[ n \right] |^2 \nonumber \\
		& + 2 \left( z_J^{\left( m \right)}\left[ n \right] - z_E\left[ n \right] \right)^T \left( z_J\left[ n \right] - z_J^{\left( m \right)}\left[ n \right] \right) \geq {D_{\min}^2} \label{Dd} \\
		& \| \mathbf{q}_J\left[ n \right] - \mathbf{q}_U\left[ n \right] \|^2 + | z_J^{\left( m \right)}\left[ n \right] - z_U^{\left( m \right)}\left[ n \right] |^2 \nonumber \\
		& + 2 \left( z_J^{\left( m \right)}\left[ n \right] - z_U\left[ n \right] \right)^T \left( z_J\left[ n \right] - z_J^{\left( m \right)}\left[ n \right] \right) \nonumber \\
		& - 2 \left( z_J\left[ n \right] - z_U^{\left( m \right)}\left[ n \right] \right)^T \left( z_U\left[ n \right] - z_U^{\left( m \right)}\left[ n \right] \right) \geq {D_{\min}^2} \label{De} \\
		& \| \mathbf{q}_U\left[ n \right] - \mathbf{q}_E\left[ n \right] \|^2 + | z_U^{\left( m \right)}\left[ n \right] - z_E\left[ n \right] |^2 \nonumber \\ 
		& + 2 \left( z_U^{\left( m \right)}\left[ n \right] - z_E\left[ n \right] \right)^T \left( z_U\left[ n \right] - z_U^{\left( m \right)}\left[ n \right] \right) \geq {D_{\min}^2}. \label{Df}
	\end{align}
\end{subequations}

Finally, by replacing $\bar{R}_{\sec}^k\left[ n \right]$ in (\ref{sp41b}) with $\bar{R}_{\sec,2}^{k, {\mathrm {lb}}}\left[ n \right]$ in (\ref{secure2lb}),
 $\mathcal{P}_{2.4 \rm{a}}$ is transformed into
\begin{subequations}
	\begin{align}
		\mathcal{P}_{2.4{\rm b}}: & \max_{\mathbf{H}, \mathbf{\Theta}, {\bf{Y}}, {\Phi _{JP}}\atop {{\bf{t}}_{P,h}}, {{\bf{p}}_h}, {\eta _k}} \sum_{k=1}^K \eta_k\\
		\mathrm { s.t. } & \frac{1}{N} \sum_{n=1}^{ N} \sum_{k=1}^K a_k\left[ n \right] \bar{R}_{\sec,2}^{k, {\mathrm {lb}}}\left[ n \right] \geq \eta_k, \label{sp4} \forall k, \\
		& \eta_k \geq R_{\min}, \forall k, \\
		&(\rm{\ref{p01e}}), (\rm{\ref{p01l}}), (\rm{\ref{p01m}}), (\rm{\ref{p01o}}), (\rm{\ref{p01p}}), \nonumber \\
		&(\rm{\ref{yySCA}}), (\rm{\ref{ITnew1}}), (\rm{\ref{A}}), (\rm{\ref{C}}), (\rm{\ref{D}}), \nonumber \\
		& (\rm{\ref{F7}}), (\rm{\ref{F8}}), (\rm{\ref{F6}}), (\rm{\ref{Dd}})-(\rm{\ref{Df}}).  \nonumber
	\end{align}
\end{subequations}
$\mathcal{P}_{2.4 \rm{b}}$ is convex and can be solved using existing optimization software such as CVX.

{\textbf{Algorithm 1} summarizes the BCD-based algorithm, which is developed to solve problem $\mathcal{P}_{1}$ sub-optimally, where $\varepsilon$ represents the accuracy of convergence.}

\subsection{Convergence Analysis of Algorithm 1}

\begin{algorithm}[t]
	\caption{Iterative Procedure of Problem $\mathcal{P}_1$}
	\KwIn{Initialization of feasible points.}
	\While{$R( \mathbf{A}^{\left( m \right)}, \mathbf{P}^{\left( m \right)}, \mathbf{\Theta}^{\left( m \right)},  \mathbf{H}^{\left( m \right)}, \mathbf{Q}^{\left( m \right)})$ - $R( \mathbf{A}^{\left( m - 1 \right)}, \mathbf{P}^{\left( m - 1 \right)},  \mathbf{\Theta}^{\left( m - 1 \right)},  \mathbf{H}^{\left( m - 1 \right)}, \mathbf{Q}^{\left( m - 1 \right)}) \succ \varepsilon$}{
		Solve $\left(\mathcal{P}_{2.1}\right)$ for given $\left\{\mathbf{P}^{\left( m \right)},\mathbf{H}^{\left( m \right)},\mathbf{Q}^{\left( m \right)},\mathbf{\Theta}^{\left( m \right)}\right\}$
		and obtain $\mathbf{A}^{\left( {m + 1} \right)}$;\\
		Solve {$\left(\mathcal{P}_{2.2 \rm{b}}\right)$} for given $\left\{\mathbf{A}^{\left( {m + 1} \right)},\mathbf{H}^{\left( m \right)},\mathbf{Q}^{\left( m \right)},\mathbf{\Theta}^{\left( m \right)}\right\}$ 
		and obtain $\mathbf{P}^{\left( {m + 1} \right)}$;\\
		Solve {$\left(\mathcal{P}_{2.3{\rm b}}\right)$} for given $\left\{\mathbf{A}^{\left( {m + 1} \right)},\mathbf{H}^{\left( m \right)},\mathbf{P}^{\left( {m + 1} \right)}\right\}$ and obtain $\mathbf{Q}^{\left( {m + 1} \right)}$;\\
		Solve {$\left(\mathcal{P}_{2.4{\rm b}}\right)$} for given $\left\{\mathbf{A}^{\left( {m + 1} \right)},\mathbf{Q}^{\left( {m + 1} \right)},\mathbf{P}^{\left( {m + 1} \right)}\right\}$ and obtain $\mathbf{H}^{\left( {m + 1} \right)}$;\\
		$m = m + 1$;\\
		Compute $R( \mathbf{A}^{\left( m \right)}, \mathbf{P}^{\left( m \right)}, \mathbf{\Theta}^{\left( m \right)},  \mathbf{H}^{\left( m \right)}, \mathbf{Q}^{\left( m \right)})$.}
	\KwOut{$R( \mathbf{A}^{\left( m \right)}, \mathbf{P}^{\left( m \right)}, \mathbf{\Theta}^{\left( m \right)},  \mathbf{H}^{\left( m \right)}, \mathbf{Q}^{\left( m \right)})$ with
		$\mathbf{A}^*=\mathbf{A}^{\left( m \right)}$, \\
		$\mathbf{P}^*=\mathbf{P}^{\left( m \right)}$,$\mathbf{\Theta}^*=\mathbf{\Theta}^{\left( m \right)}$, $\mathbf{H}^*=\mathbf{H}^{\left( m \right)}$,$\mathbf{Q}^*=\mathbf{Q}^{\left( m \right)}$.}
\end{algorithm}

In \textbf{Algorithm 1}, step 1 addresses the user scheduling problem $\mathcal{P}_{2.1}$ by optimizing for given feasible variables $\mathbf{P}^{(m)}$, $\mathbf{H}^{(m)}$, $\mathbf{Q}^{(m)}$, and $\mathbf{\Theta}^{(m)}$. This results in
\begin{align}\label{z1.1}
	&R\left(\mathbf{A}^{(m)}, \mathbf{P}^{(m)}, \mathbf{\Theta}^{(m)}, \mathbf{H}^{(m)}, \mathbf{Q}^{(m)}\right) \nonumber\\
	&\leq R\left(\mathbf{A}^{\left( {m + 1} \right)}, \mathbf{P}^{(m)}, \mathbf{\Theta}^{(m)}, \mathbf{H}^{(m)}, \mathbf{Q}^{(m)}\right).
\end{align}

In step 2, we solve the power optimization problem $\mathcal{P}_{2.2 \rm{b}}$ with fixed variables $\mathbf{A}^{\left( {m + 1} \right)}$, $\mathbf{H}^{(m)}$, $\mathbf{Q}^{(m)}$, and $\mathbf{\Theta}^{(m)}$. Since this problem is convex, we obtain
\begin{align}\label{z2.2}
	&R\left(\mathbf{A}^{\left( {m + 1} \right)}, \mathbf{P}^{(m)}, \mathbf{\Theta}^{(m)}, \mathbf{H}^{(m)}, \mathbf{Q}^{(m)}\right) \nonumber \\
	&\leq R\left(\mathbf{A}^{\left( {m + 1} \right)}, \mathbf{P}^{\left( {m + 1} \right)}, \mathbf{\Theta}^{(m)}, \mathbf{H}^{(m)}, \mathbf{Q}^{(m)}\right).
\end{align}

Step 3 involves solving the approximate horizontal trajectory optimization problem $\mathcal{P}_{2.3 \rm{b}}$ by fixing $\mathbf{A}^{\left( {m + 1} \right)}$, $\mathbf{P}^{\left( {m + 1} \right)}$, $\mathbf{\Theta}^{(m)}$, and $\mathbf{H}^{(m)}$, leading to
\begin{align}\label{z3.2}
	&R\left(\mathbf{A}^{\left( {m + 1} \right)}, \mathbf{P}^{\left( {m + 1} \right)}, \mathbf{\Theta}^{(m)}, \mathbf{H}^{(m)}, \mathbf{Q}^{(m)}\right) \nonumber\\
	& \stackrel{(a)}{=} R^{\mathrm{lb}}\left(\mathbf{A}^{\left( {m + 1} \right)}, \mathbf{P}^{\left( {m + 1} \right)}, \mathbf{\Theta}^{(m)}, \mathbf{H}^{(m)}, \mathbf{Q}^{(m)}\right) \nonumber\\
	& \stackrel{(b)}{\leq} R^{\mathrm{lb}}\left(\mathbf{A}^{\left( {m + 1} \right)}, \mathbf{P}^{\left( {m + 1} \right)}, \mathbf{\Theta}^{(m + 1)}, \mathbf{H}^{(m)}, \mathbf{Q}^{\left( {m + 1} \right)}\right) \nonumber\\
	& \stackrel{(c)}{\leq} R\left(\mathbf{A}^{\left( {m + 1} \right)}, \mathbf{P}^{\left( {m + 1} \right)}, \mathbf{\Theta}^{(m + 1)}, \mathbf{H}^{(m)}, \mathbf{Q}^{\left( {m + 1} \right)}\right).
\end{align}
Here, 
(a) holds as $\mathcal{P}_{2.3 \rm{b}}$ yield the same objective for $\mathbf{Q}^{(m)}$, 
(b) is true since $\mathbf{Q}^{\left( {m + 1} \right)}$ maximizes the lower bound $R^{\mathrm{lb}}$ while keeping $\mathbf{A}^{\left( {m + 1} \right)}$, $\mathbf{H}^{(m)}$, and $\mathbf{P}^{\left( {m + 1} \right)}$ fixed, 
(c) follows as the target value of $\mathcal{P}_{2.3 \rm{b}}$ serves as a lower bound for the original problem at $\mathbf{Q}^{\left( {m + 1} \right)}$. Thus, the objective value of $\mathcal{P}_{2.3 \rm{b}}$ does not decrease with iterations, even with an approximate UAV trajectory.

In step 4, we fix $\mathbf{A}^{\left( {m + 1} \right)}$, $\mathbf{Q}^{\left( {m + 1} \right)}$, and $\mathbf{P}^{\left( {m + 1} \right)}$ to achieve
\begin{align}\label{z4.2}
	&R\left(\mathbf{A}^{\left( {m + 1} \right)}, \mathbf{P}^{\left( {m + 1} \right)}, \mathbf{\Theta}^{\left( m + 1 \right)}, \mathbf{H}^{\left( m \right)}, \mathbf{Q}^{\left( {m + 1} \right)}\right) \nonumber\\
	& \stackrel{(a)}{=} R^{\mathrm{lb}}\left(\mathbf{A}^{\left( {m + 1} \right)}, \mathbf{P}^{\left( {m + 1} \right)}, \mathbf{\Theta}^{\left( m+1 \right)}, \mathbf{H}^{\left( m \right)}, \mathbf{Q}^{\left( {m + 1} \right)}\right) \nonumber\\
	& \stackrel{(b)}{\leq} R^{\mathrm{lb}}\left(\mathbf{A}^{\left( {m + 1} \right)}, \mathbf{P}^{\left( {m + 1} \right)}, \boldsymbol{\Theta}^{\left( {m + 1} \right)}, \mathbf{H}^{\left( {m + 1} \right)}, \mathbf{Q}^{\left( {m + 1} \right)}\right) \nonumber\\
	& \stackrel{(c)}{\leq} R\left(\mathbf{A}^{\left( {m + 1} \right)}, \mathbf{P}^{\left( {m + 1} \right)}, \mathbf{\Theta}^{\left( {m + 1} \right)}, \mathbf{H}^{\left( {m + 1} \right)}, \mathbf{Q}^{\left( {m + 1} \right)}\right).
\end{align}
This is similar to the representation in (\ref{z3.2}), and from (\ref{z1.1}) to (\ref{z4.2}), we obtain
\begin{align}\label{z4.3}
	&R\left(\mathbf{A}^{\left( m \right)}, \mathbf{P}^{\left( m \right)}, \mathbf{\Theta}^{\left( m \right)}, \mathbf{H}^{\left( m \right)}, \mathbf{Q}^{\left( m \right)}\right) \nonumber\\
	&\leq R\left(\mathbf{A}^{\left( {m + 1} \right)}, \mathbf{P}^{\left( {m + 1} \right)}, \mathbf{\Theta}^{\left( {m + 1} \right)}, \mathbf{H}^{\left( {m + 1} \right)}, \mathbf{Q}^{\left( {m + 1} \right)}\right).
\end{align}
The analysis shows that the objective value of $\mathcal{P}_1$ does not decrease with each iteration of \textbf{Algorithm 1}, ensuring convergence due to its finite upper bound. Simulation results demonstrate that the BCD-based method converges quickly under the given conditions.

Each iteration involves solving convex optimization problems with polynomial complexity, allowing efficient convergence even in wireless networks with a moderate number of GDs. By iteratively solving the subproblems $\mathcal{P}_{2.1}$, $\mathcal{P}_{2.2 \rm{b}}$, $\mathcal{P}_{2.3 \rm{b}}$, and $\mathcal{P}_{2.4 \rm{b}}$ until achieving the desired accuracy $\varepsilon > 0$, assuming that the number of iterations for the inner and outer loops of \textbf{Algorithm 1} are denoted by $ M_2 $ and $ M_1 $, respectively. We obtain a suboptimal solution for $\mathcal{P}_1$ with a computational complexity of $O\left(M_1\left(\sqrt{N(K+1)} \log \frac{1}{\varepsilon}+M_2((K+1) N)^3 \log \frac{1}{\varepsilon}\right)\right)$ \cite{LiM2022IOT}.

\color{black}

\begin{table}
	\centering
	\caption{Simulation Parameters.}
	\begin{threeparttable}
		\resizebox{0.45\textwidth}{!}
		{
			\begin{tabular}{ c | c| c| c}
				\Xhline{1.2pt}
				\textbf{Parameter}   				&\textbf{Value} 	&\textbf{Parameter}   				&\textbf{Value} 	\\
				\hline
				$\left[ {{{\bf{q}}_U^{\mathrm I}},{z_U^{\mathrm I}}} \right]$& ${\left[ {{\mathrm{-200,0, 30}}} \right]^T}$
				&$\left[ {{{\bf{q}}_U^{\mathrm F}},{z_U^{\mathrm F}}} \right]$& ${\left[ {{\mathrm{200,0, 30}}} \right]^T}$ \\
				\hline
				$\left[ {{{\bf{q}}_J^{\mathrm I}},{z_J^{\mathrm I}}} \right]$& ${\left[ {{\mathrm{-200,0, 50}}} \right]^T}$
				&$\left[ {{{\bf{q}}_J^{\mathrm F}},{z_J^{\mathrm F}}} \right]$& ${\left[ {{\mathrm{200,0, 50}}} \right]^T}$ \\
				\hline
				$\left[ {{{\bf{q}}_E^{\mathrm I}},{z_E^{\mathrm I}}} \right]$& ${\left[ {{\mathrm{-250,150, 85}}} \right]^T}$
				&$\left[ {{{\bf{q}}_E^{\mathrm F}},{z_E^{\mathrm F}}} \right]$& ${\left[ {{\mathrm{250,-150, 85}}} \right]^T}$ \\
				\hline
				${{\mathbf{w}}_{k}}$             &$\left[ {_{50, - 60,60, - 50}^{ - 150, - 90,60,160}} \right]$
				&$H_{\min}, H_{\max}$  				&30 m, 150 m\\
				\hline
				${\sigma ^2}$						&$-100$ dBm
				&${P_K^{ave}}$, ${P_K^{\max} }$	    &0.05 W, 0.2 W\\
				\hline
				${P_J^{ave}}$, ${P_J^{\max}}$      &0.025 W, 0.1 W
				&${P_{{U,\mathrm {ave }}}^{{\mathrm{hor}}}}$   & 175 W \\
				\hline
				${P_{{U,\mathrm {ave }}}^{{\mathrm{ver}}}}$   & 65 W 
				&${P_{{J,\mathrm {ave }}}^{{\mathrm{hor}}}}$   & 130 W \\
				\hline
				${P_{{J,\mathrm {ave }}}^{{\mathrm{ver}}}}$   & 30 W 
				&${\alpha_{{\mathrm L}}}$, ${\alpha_{{\mathrm N}}}$			&2.2, 3.4 \\
				\hline
				${a}, {b}$			                &11.95, 0.14
				&$V_{xy}, V_{z}$						&25 m/s, 10 m/s\\
				\hline
				${\rho _0}$, ${\mu}$				    	&$-60$ dB, $-20$ dB
				&$R_{\min}$					        	&0.3 bits/Hz\\
				\hline
				${D_{\min}}$, ${r_E}$					    &15 m, 10m
				&${T}$					        	&40 s\\
				\hline
				${\delta _t}$						& 1 s
				&$\Gamma$						& $-105$ dB\\
				\Xhline{1.2pt}
			\end{tabular}}
		\end{threeparttable}
	\label{table3}
\end{table}

\begin{figure*}[t]
	\centering
	\subfigure[ASSE with different schemes.]{
		\label{fig02a}
		\includegraphics[width = 0.23 \textwidth]{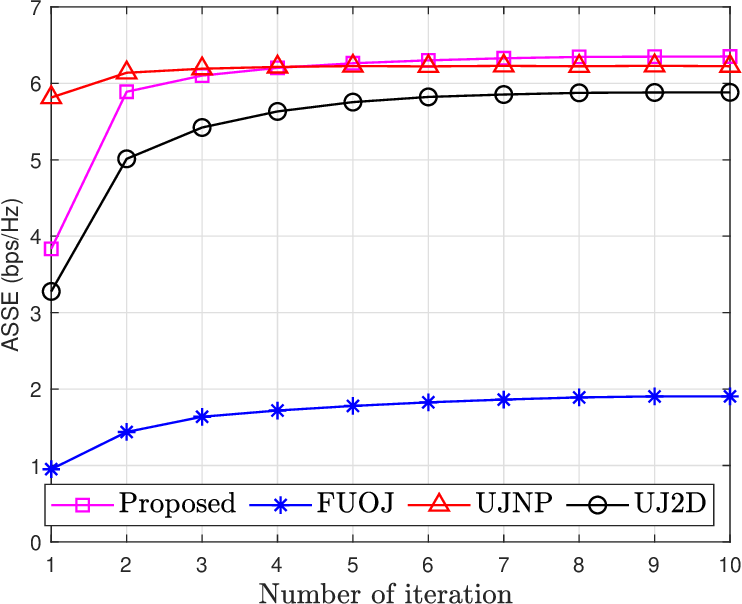}}
	\subfigure[ASSE of the proposed scheme with varying power budget.]{
		\label{fig02b}
		\includegraphics[width = 0.23 \textwidth]{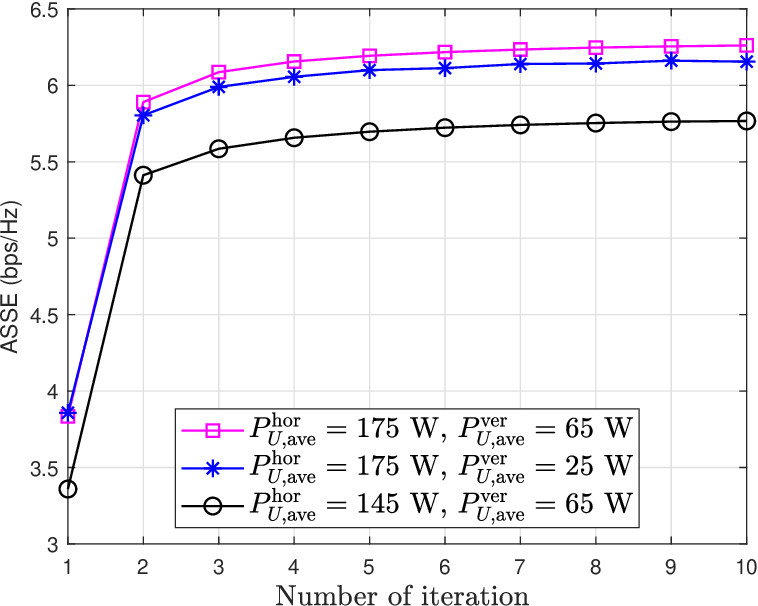}}
	\subfigure[ASSE of the proposed scheme with varying $\Gamma$.]{
		\label{fig02c}
		\includegraphics[width = 0.23 \textwidth]{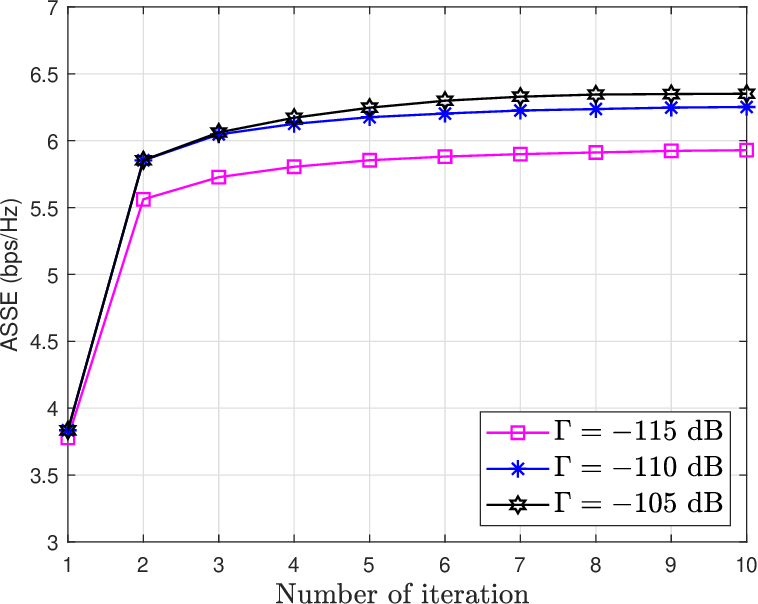}}	
	\subfigure[ASSE of the proposed scheme with varying $T$.]{
		\label{fig02d}
		\includegraphics[width = 0.23 \textwidth]{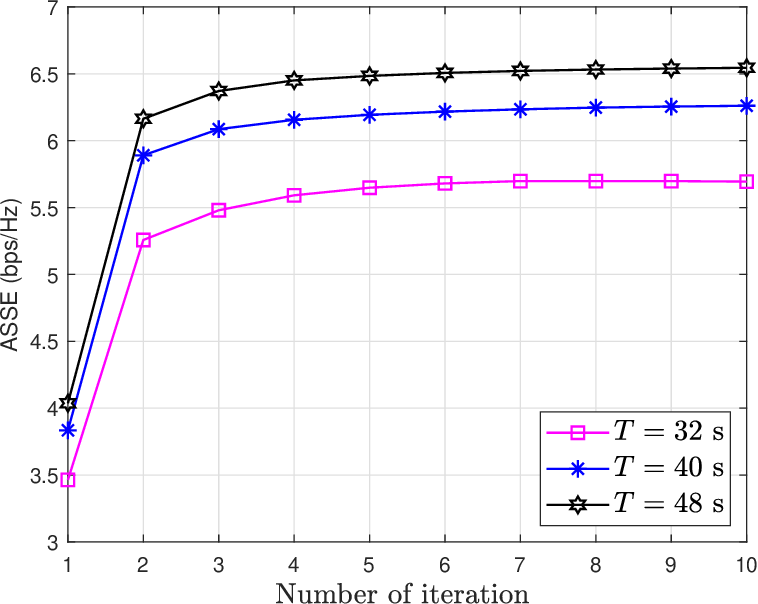}}	
	\caption{ASSE versus the number of iterations.}
	\label{fig02}
\end{figure*}

\begin{figure*}[t]
	\centering
	\subfigure[2D trajectories of UAVs with the proposed scheme.]{
		\label{fig03a}
		\includegraphics[width = 0.23  \textwidth]{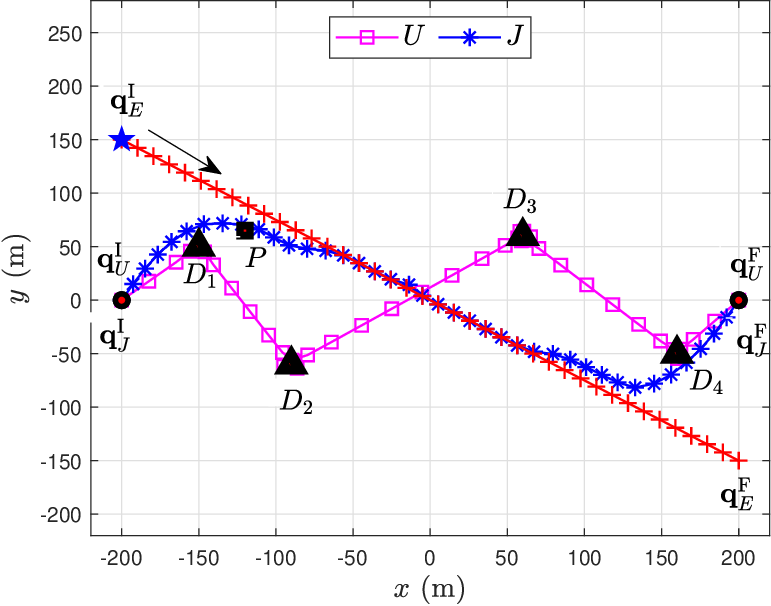}}
	\subfigure[2D trajectories of UAVs with FUOJ scheme.]{
		\label{fig03b}
		\includegraphics[width = 0.23  \textwidth]{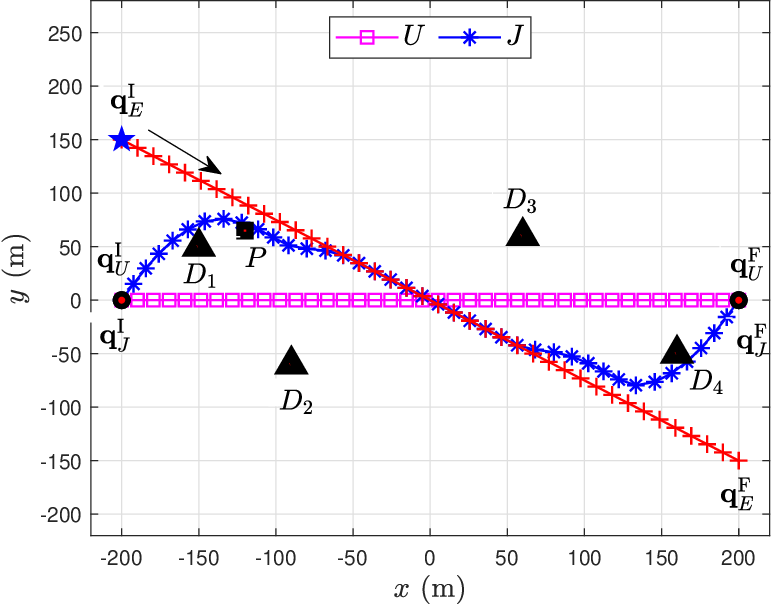}}	
	\subfigure[2D trajectories of UAVs with UJNP scheme.]{
		\label{fig03c}
		\includegraphics[width = 0.23  \textwidth]{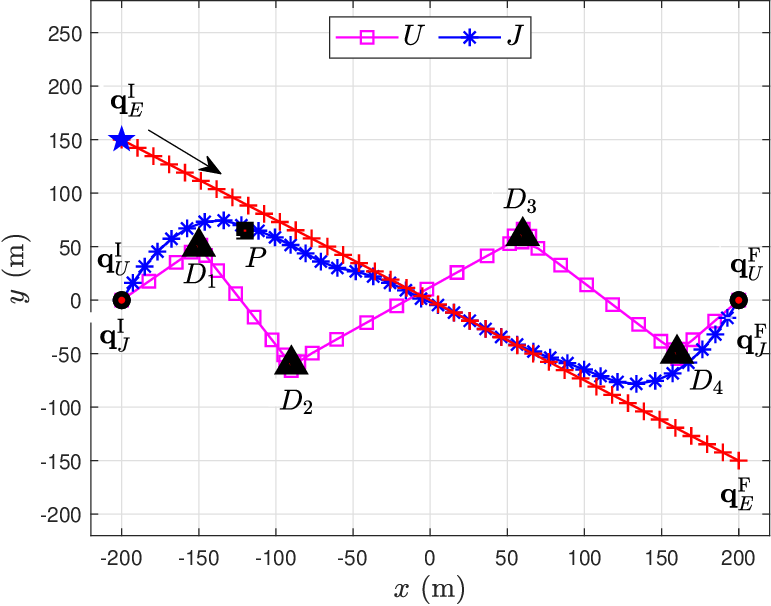}}
	\subfigure[2D trajectories of UAVs with UJ2D scheme.]{
		\label{fig03d}
	\includegraphics[width = 0.23  \textwidth]{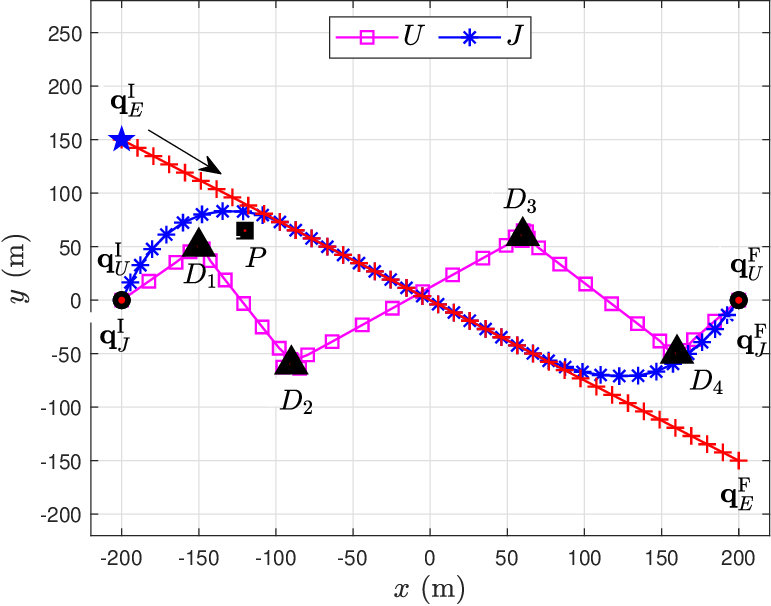}}
	\subfigure[3D trajectories of UAVs with the proposed scheme.]{
		\label{fig03e}
		\includegraphics[width = 0.23  \textwidth]{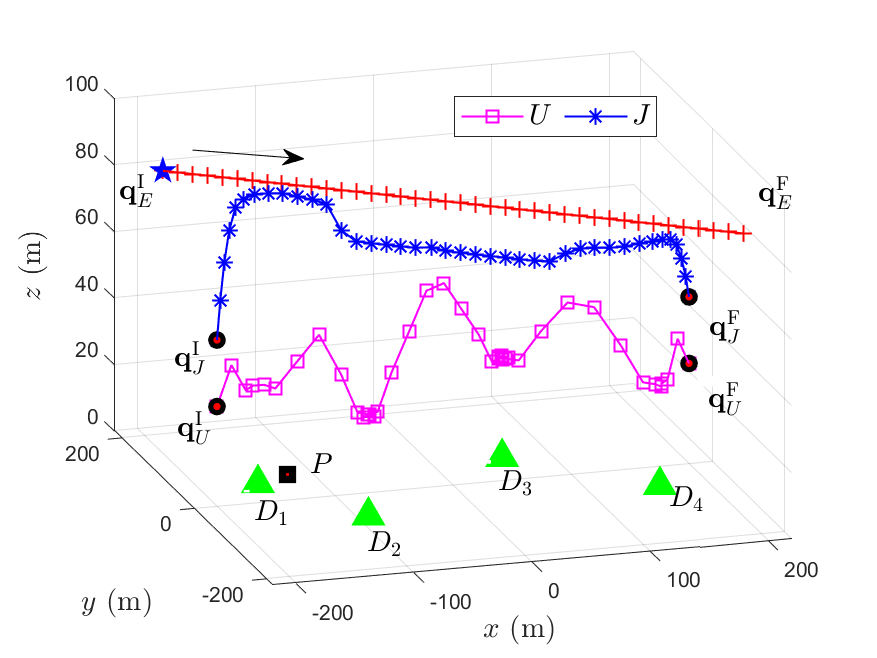}}	
	\subfigure[3D trajectories of UAVs with FUOJ scheme.]{
		\label{fig03f}
		\includegraphics[width = 0.23  \textwidth]{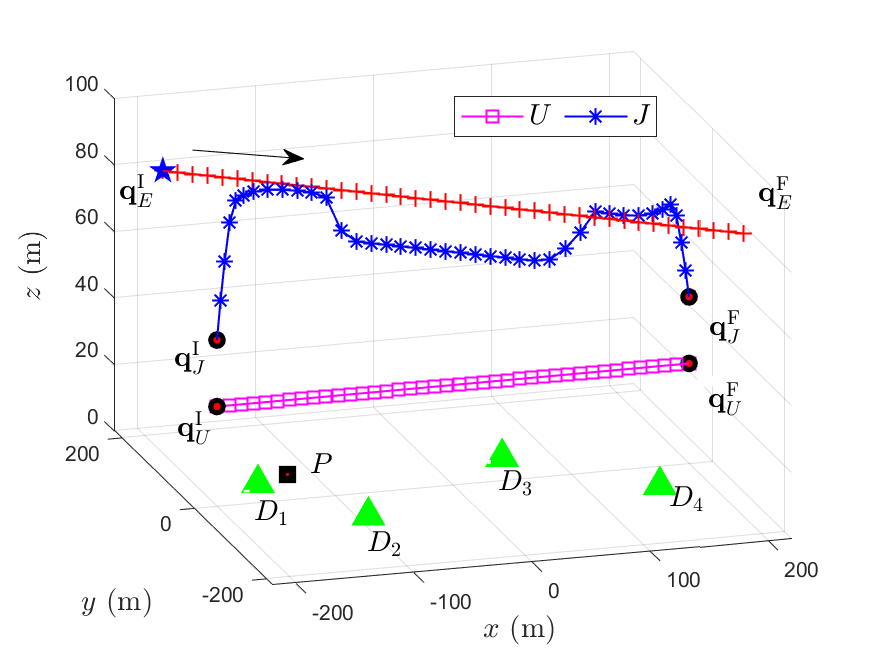}}
	\subfigure[3D trajectories of UAVs with UJNP scheme.]{
		\label{fig03g}
		\includegraphics[width = 0.23  \textwidth]{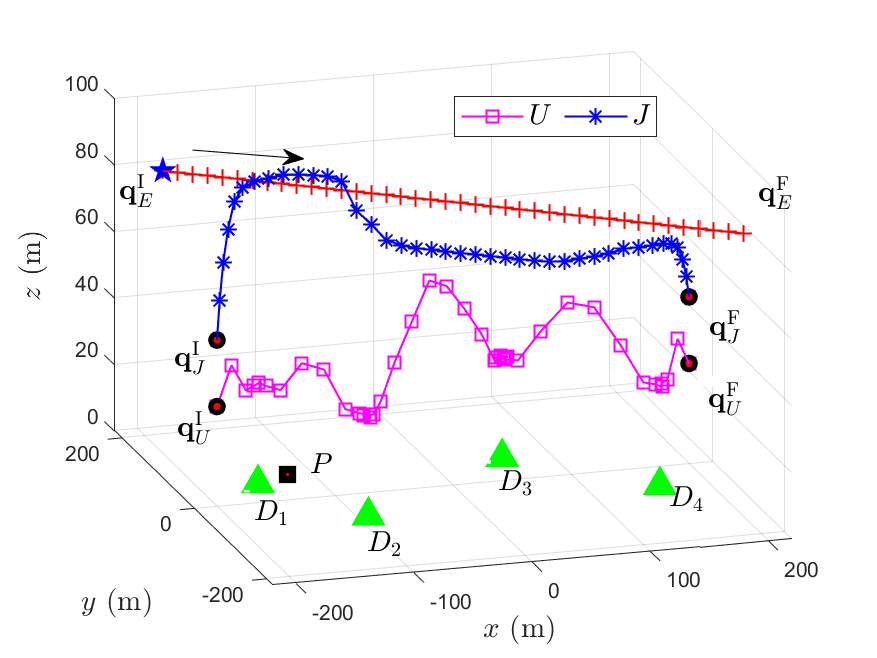}}
	\subfigure[3D trajectories of UAVs with UJ2D scheme.]{
		\label{fig03h}
		\includegraphics[width = 0.23  \textwidth]{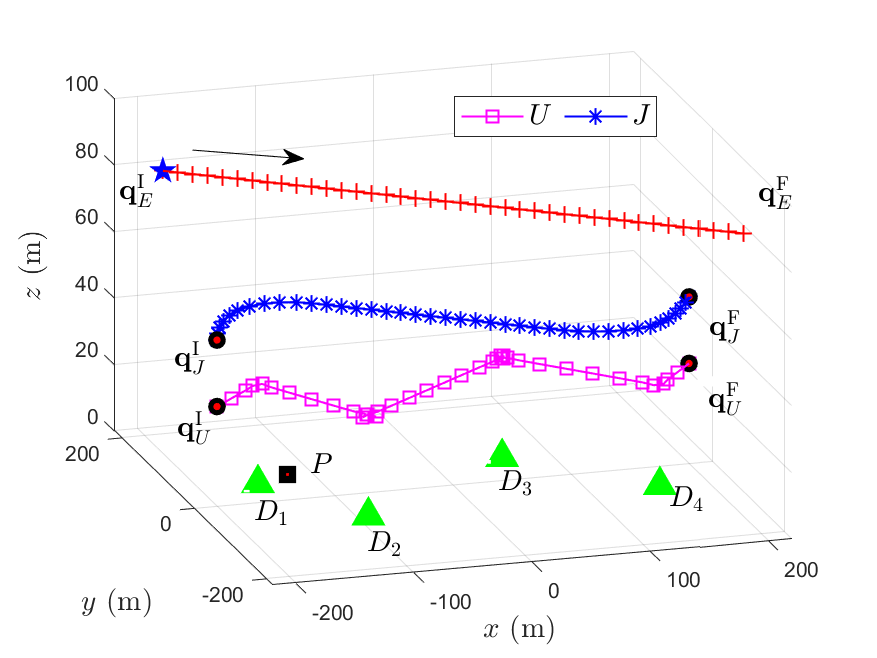}}		
	\caption{{Trajectories of UAVs with different schemes.}}
	\label{fig03}
\end{figure*}

\begin{figure*}[t]
	\centering
	\subfigure[Transmit power with the proposed scheme.]{
		\label{fig04a}
		\includegraphics[width = 0.23 \textwidth]{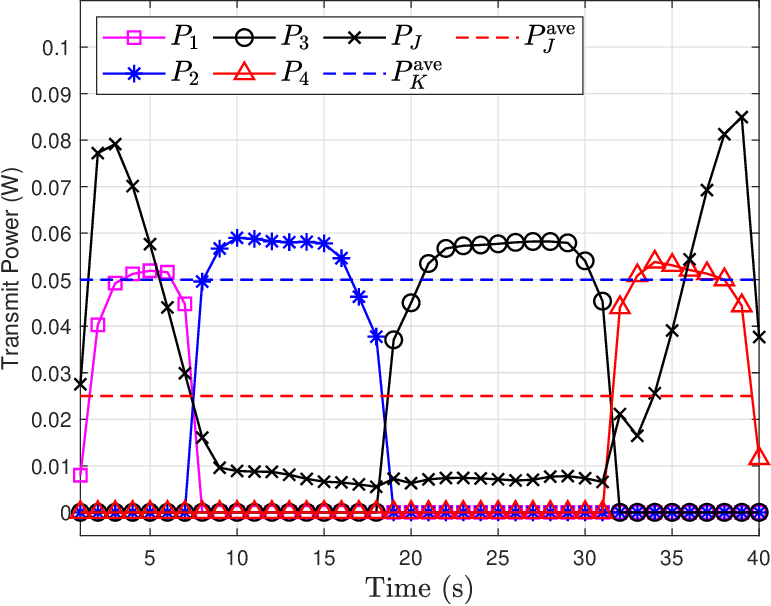}}
	\subfigure[Transmit power with the FUOJ scheme.]{
		\label{fig04b}
		\includegraphics[width = 0.23 \textwidth]{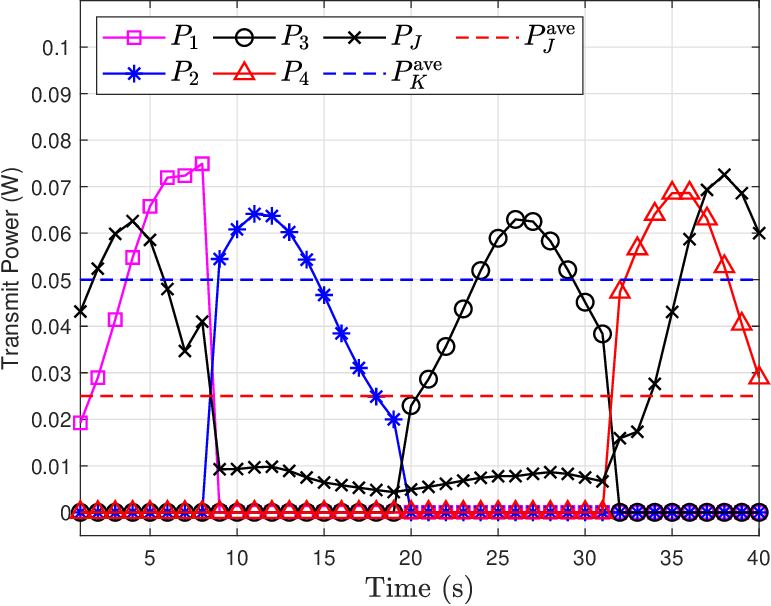}}
	\subfigure[Transmit power with the UJ2D scheme]{
		\label{fig04c}
		\includegraphics[width = 0.23 \textwidth]{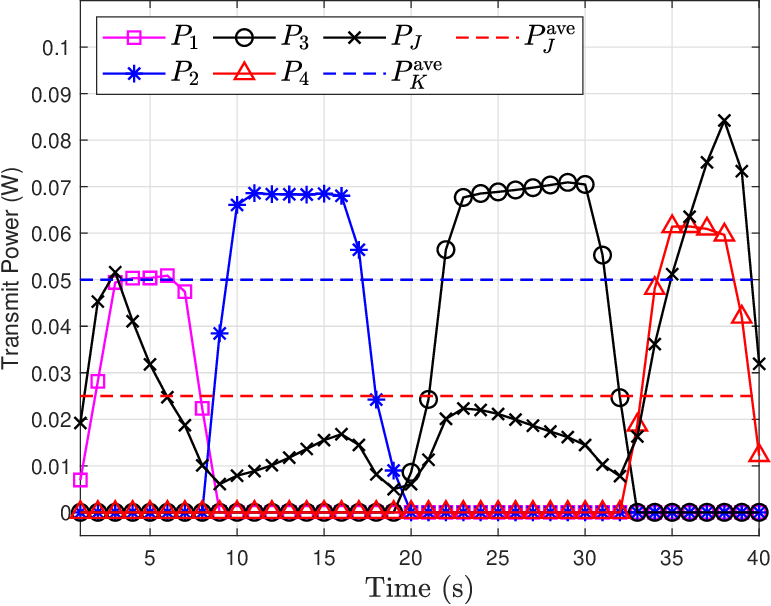}}
	\subfigure[SSE with different schemes.]{
		\label{fig04d}
		\includegraphics[width = 0.23 \textwidth]{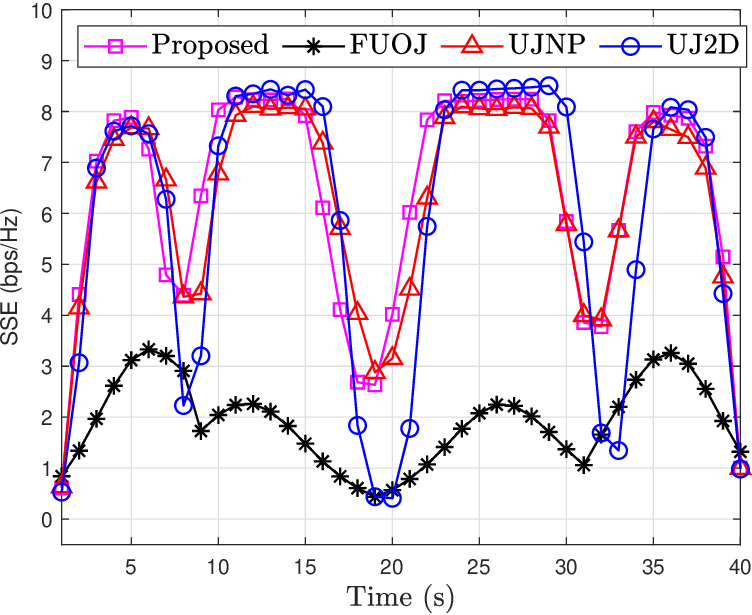}}	
	\caption{{Transmit power of different schemes during the flight time.}}
	\label{fig04}
\end{figure*}

\begin{figure*}[t]
	\centering
	\subfigure[2D trajectory of $U$.]{
		\label{fig05a}
		\includegraphics[width = 0.23 \textwidth]{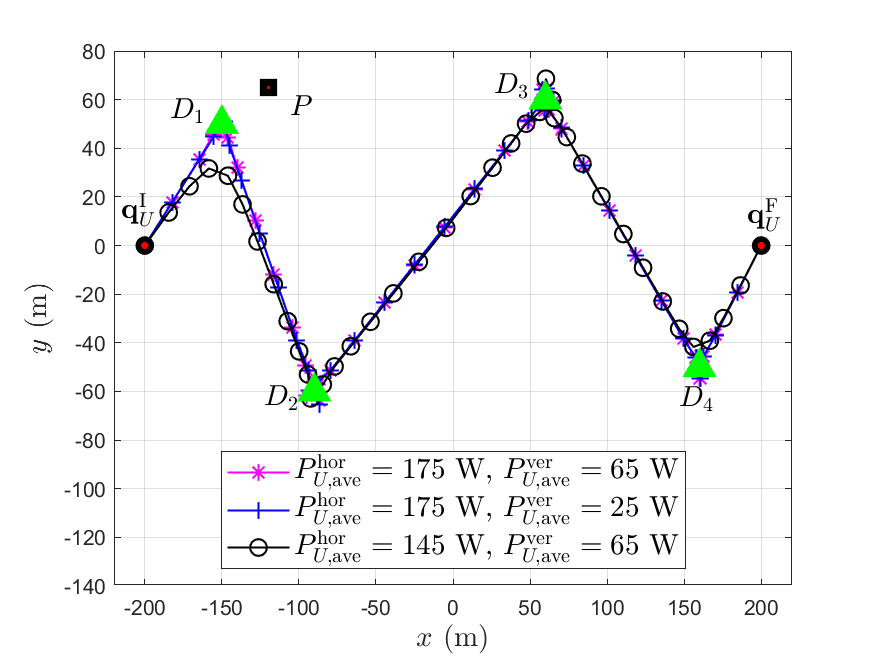}}			
	\subfigure[3D trajectory of $U$.]{
		\label{fig05b}
		\includegraphics[width = 0.23 \textwidth]{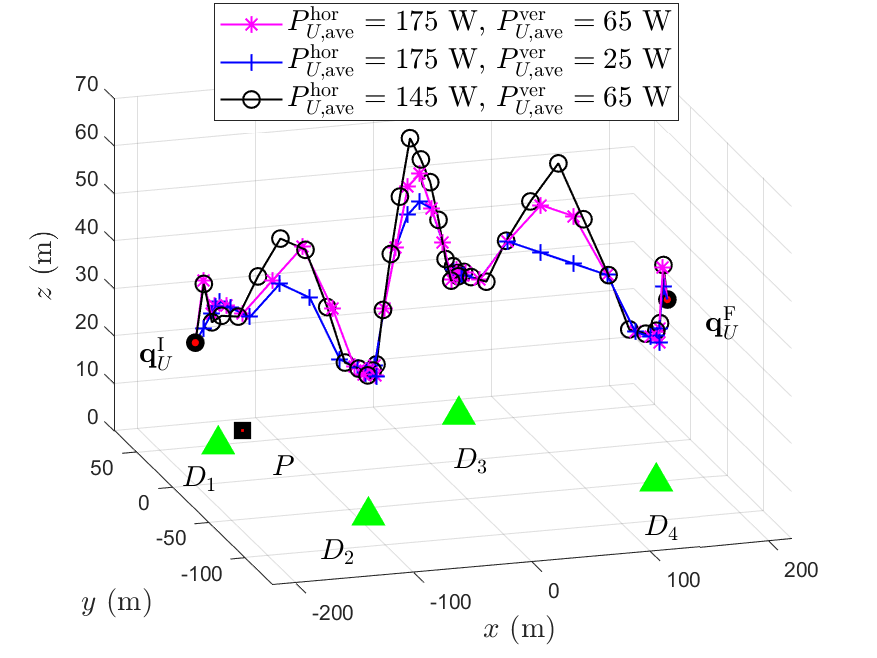}}
	\subfigure[Horizontal velocity of $U$.]{
		\label{fig05c}
		\includegraphics[width = 0.23 \textwidth]{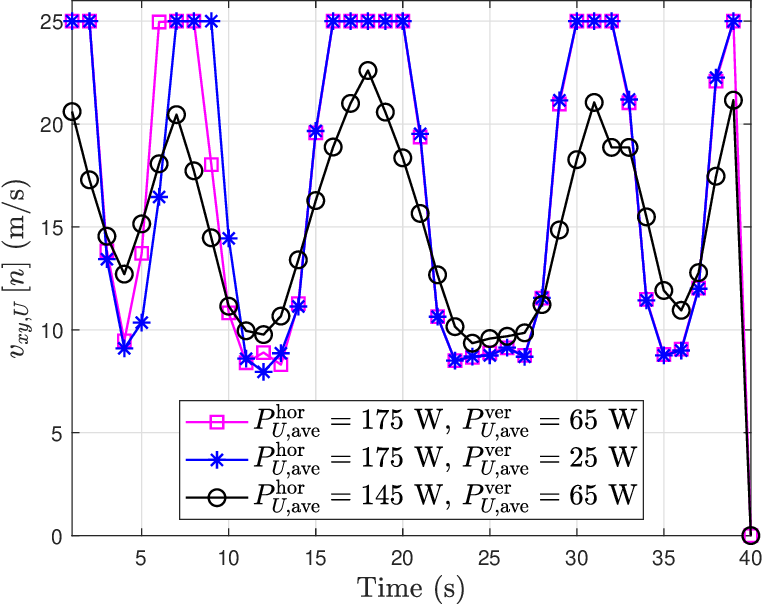}}			
	\subfigure[Vertical velocity of $U$.]{
		\label{fig05d}
		\includegraphics[width = 0.23 \textwidth]{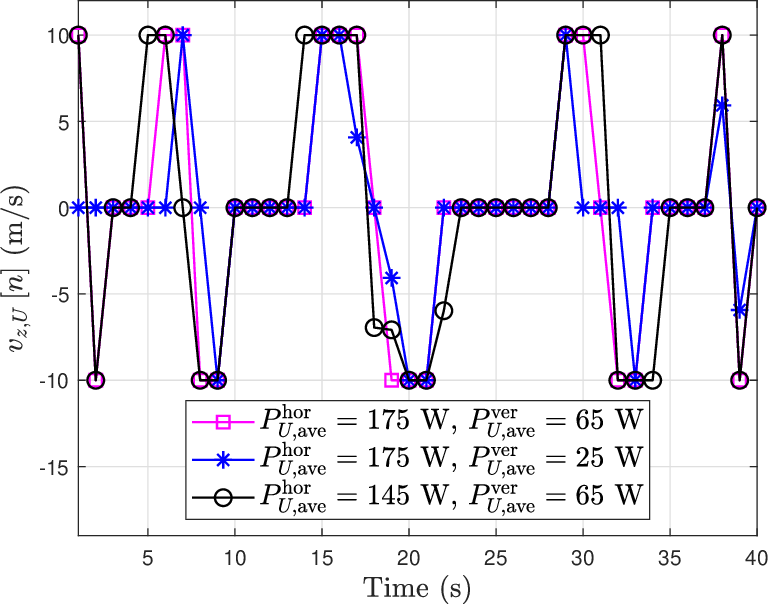}}
	\caption{{Trajectory and velocity of $U$.}}
	\label{fig05}
\end{figure*}

\begin{figure*}[t]
	\centering
	\subfigure[2D trajectory of $J$.]{
		\label{fig06a}
		\includegraphics[width = 0.23 \textwidth]{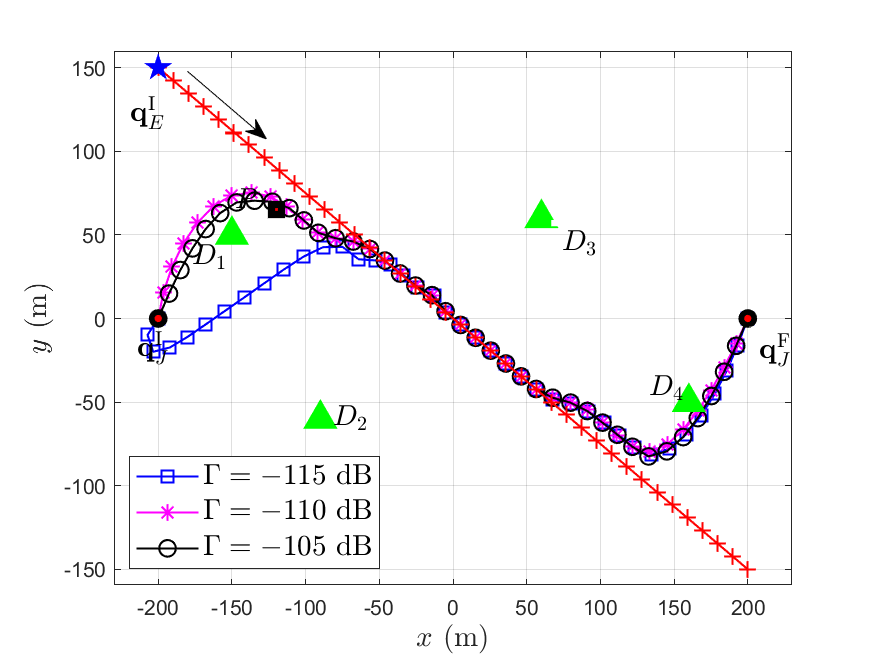}}
	\subfigure[3D trajectory of $J$.]{
		\label{fig06b}
		\includegraphics[width = 0.23 \textwidth]{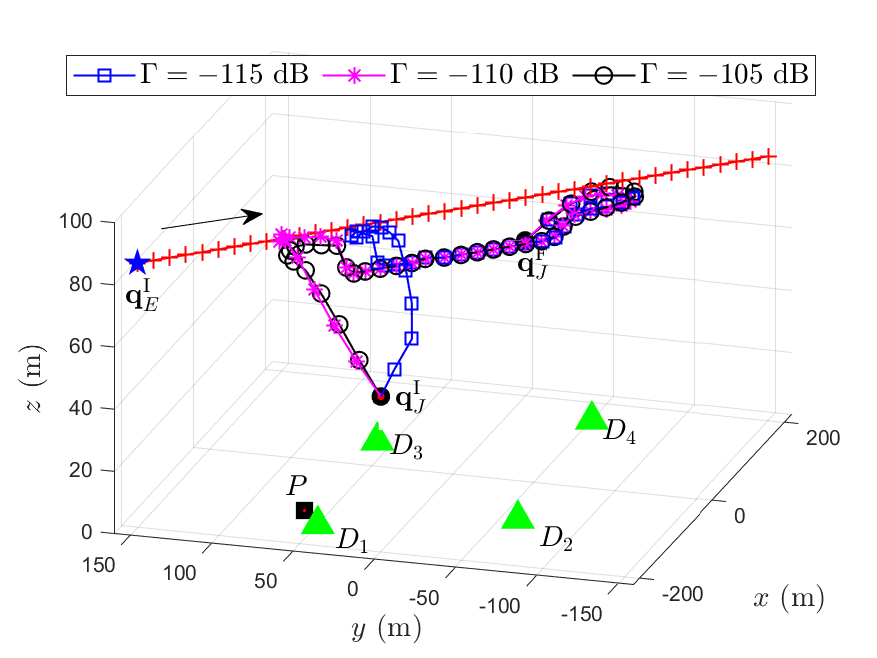}}
	\subfigure[Transmit power of $J$.]{
		\label{fig06c}
		\includegraphics[width = 0.23 \textwidth]{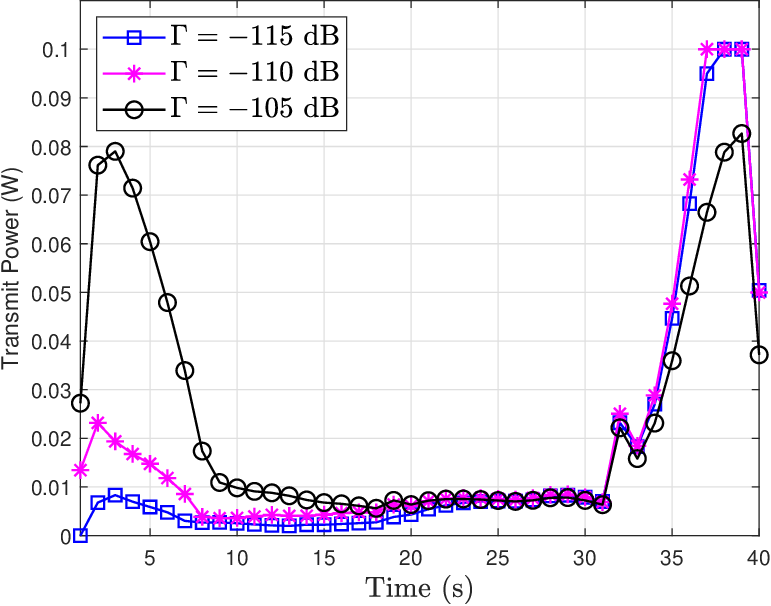}}
	\subfigure[SSE versus varying $\Gamma$.]{
		\label{fig06d}
		\includegraphics[width = 0.23 \textwidth]{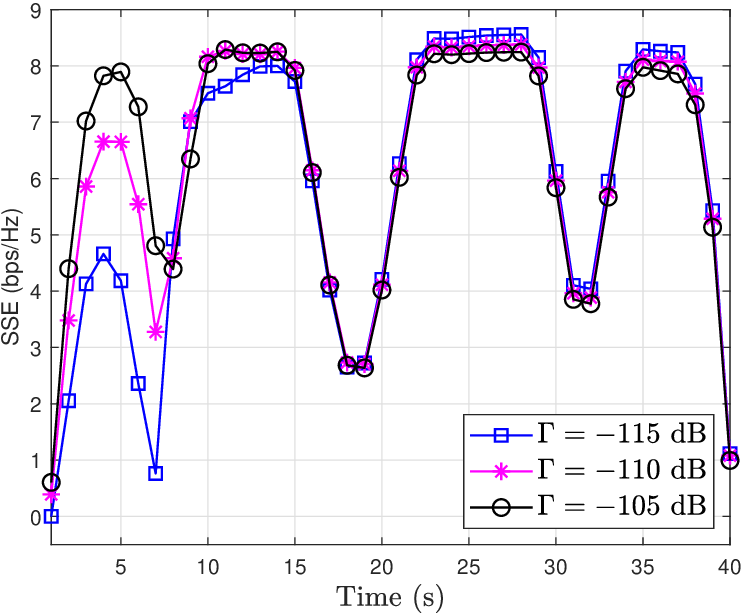}}	
	\caption{Trajectory and transmit power of $J$ and SSE versus varying $\Gamma$.}
	\label{fig06}
\end{figure*}

\begin{figure*}[t]
	\centering
	\subfigure[The ASSE versus $ T$.]{
		\label{fig07a}
		\includegraphics[width = 0.23 \textwidth]{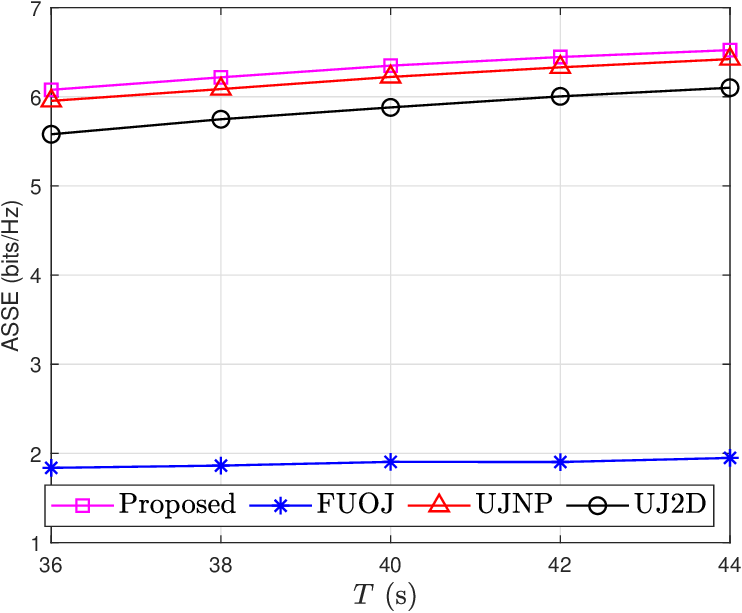}}			
	\subfigure[The ASSE versus ${P_{{U,\mathrm {ave }}}^{{\mathrm{hor}}}}$.]{
		\label{fig07b}
		\includegraphics[width = 0.23 \textwidth]{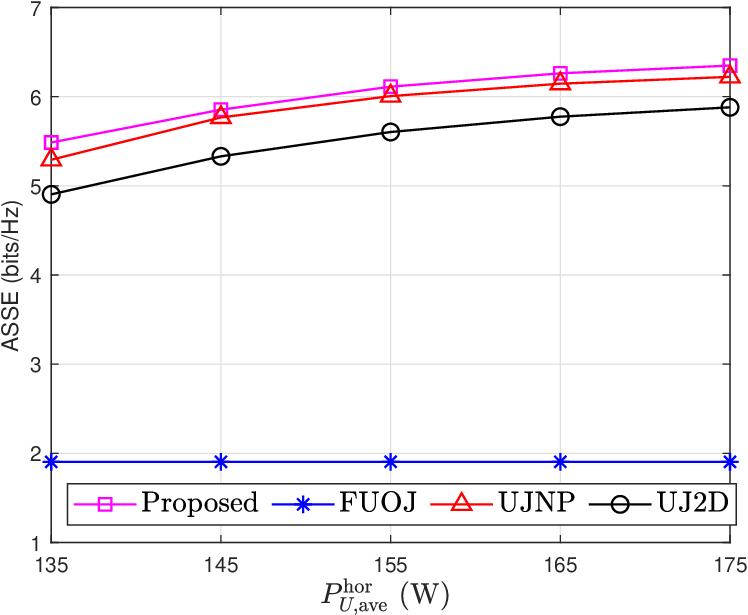}}
	\subfigure[The ASSE versus ${r_{E}}$.]{
		\label{fig07c}
		\includegraphics[width = 0.23 \textwidth]{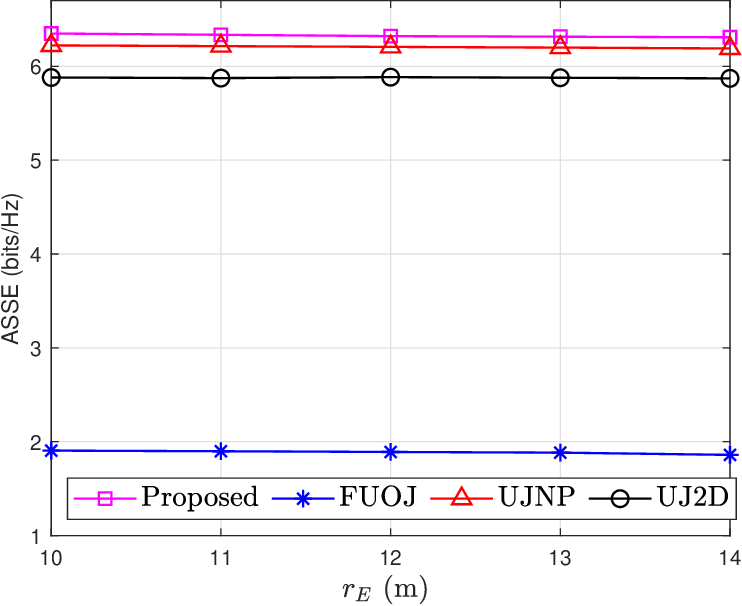}}			
	\subfigure[The ASSE versus $\Gamma$.]{
		\label{fig07d}
		\includegraphics[width = 0.23 \textwidth]{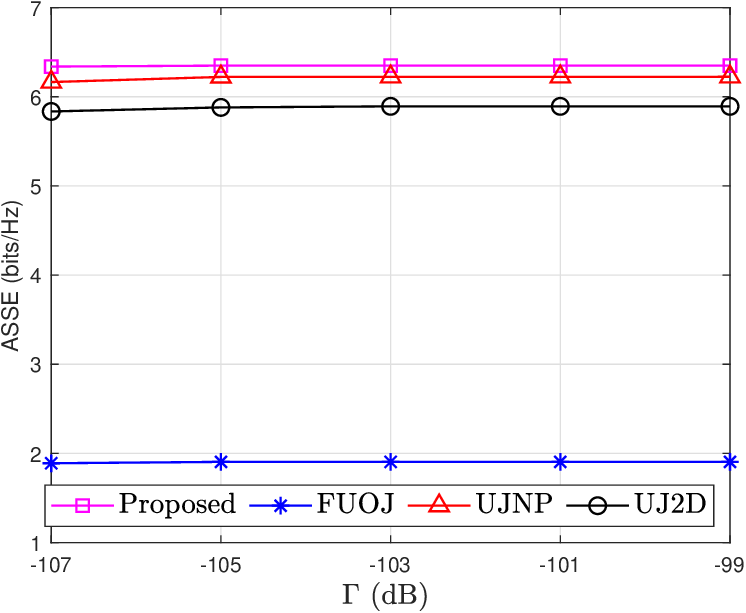}}
	\caption{{ASSE with different schemes.}}
	\label{fig07}
\end{figure*}

\section{Numerical Results}
\label{Simulation}

In this section, we present numerical results to demonstrate the effectiveness of the proposed algorithm. Without loss of generality, {four GDs are distributed within a square area of $400 \, \mathrm{m} \times 400 \, \mathrm{m}$. }
All parameters for the system under consideration are detailed in Table \ref{table3} \cite{LiA2019WCL}, \cite{LiuZ2024TVT}, \cite{LeiH2023IoT}, \cite{DuoB2020TVT3D}, \cite{YouC2020TWC}. Additionally, the parameters in (\ref{speedxy}) are configured based on the example provided in \cite{ZhangR2021TWC}. 
To highlight the benefits of the proposed scheme, we compare it against the following three benchmarks.
\begin{enumerate}		
	\item  Benchmark I (denoted by `FUOJ'): 
	Similar to \cite{NguyenX2021TVT}, the power and trajectory of $J$, the transmit power of GDs, and user scheduling are jointly optimized, with the fixed trajectory of $U$. {It aims to quantify the additional performance gain achieved by introducing \textit{optimization of 3D trajectory for $U$}.	}
	
	\item  {Benchmark II (denoted by `UJNP'): In this scheme, the 3D trajectory of $U$ and $J$, and user scheduling are jointly optimized, with the fixed power of GDs and $J$, the fixed power levels are their average values ${P_K^{ave}}$ and ${P_J^{ave}}$. {It aims to quantify the additional performance gain achieved by introducing \textit{power optimization for GDs and $J$}.	}}	
	
	\item Benchmark III (denoted by `UJ2D'): 
	Similar to \cite{LeiH2024TCCNFair}, all UAVs operate at the minimum altitude, the 2D trajectories of $U$ and $J$, the jamming power from $J$, the transmit power of GDs, and user scheduling are jointly designed. {It aims to quantify the additional performance gain achieved by introducing \textit{optimization of vertical trajectory for $U$ and  $J$}.}	
\end{enumerate}

Fig. \ref{fig02a} plots the convergence of all the schemes, and the curves demonstrate that the proposed scheme exhibits superior convergence performance and achieves the highest secrecy rate compared to benchmarks, which underscores the effectiveness and benefits of the proposed algorithm. 
{ Compared with the benchmark schemes, the proposed scheme demonstrates performance advantages by jointly optimizing the 3D trajectories of dual UAVs: the FUOJ scheme cannot flexibly approach GDs due to the lack of optimized communication UAV trajectory; The UJ2D scheme, due to its fixed height, limits the ability of communication UAVs to obtain better elevation angles and interfere with the ability of UAVs to approach eavesdropper. This collectively reflects the crucial role of 3D trajectory design in improving the overall performance of the system. Compared with the two benchmark schemes mentioned above, the importance of trajectory design is demonstrated. At the same time, we compared the UJNP scheme with fixed power optimized trajectory, and its performance is still lower than the proposed scheme. Fixed power will reduce performance, especially in scenarios such as CRN that require power control.}
Figs. \ref{fig02b}-\ref{fig02d} show the convergence behavior of the proposed scheme with varying the propulsion power budgets of $U$ (${P_{{U,\mathrm {ave }}}^{{\mathrm{hor}}}}$ and ${P_{{U,\mathrm {ave }}}^{{\mathrm{ver}}}}$), the IT threshold ($\Gamma$), and the flight duration ($T$). 
It can be observed that increasing ${P_{{U,\mathrm {ave }}}^{{\mathrm{hor}}}}$ and ${P_{{U,\mathrm {ave }}}^{{\mathrm{ver}}}}$ can enhance the secrecy performance since more power budget can obtain better channel quality. 
{As shown in Fig. \ref{fig02b}, when ${P_{{U,\mathrm {ave }}}^{{\mathrm{hor}}}}$ = 175 W and ${P_{{U,\mathrm {ave }}}^{{\mathrm{ver}}}}$ = 65 W, the system's ASSE is 6.252 bps/Hz, while in the case of ${P_{{U,\mathrm {ave }}}^{{\mathrm{ver}}}}$ = 25 W, the ASSE is 6.161 bps/Hz, with a gain of 1.48\%. When ${P_{{U,\mathrm {ave }}}^{{\mathrm{hor}}}}$ = 145 W, the ASSE in this case is 5.762 bps/Hz, and the gain compared to this case is 8.5\%, indicating the importance of horizontal energy consumption. In situations where energy consumption is limited, horizontal energy consumption should be considered first.}
The results in Fig. \ref{fig02c} prove that the ASSE with larger $\Gamma$ outperforms that with lower $\Gamma$ since a small $\Gamma$ denotes a tighter constraint for the primary users. 
{It can be observed that the proposed scheme has a system ASSE of 5.923 bps/Hz at $\Gamma $ = -115 dB. As $\Gamma$ increases, the corresponding ASSE are 6.247 bps/Hz and 6.348 bps/Hz, respectively, resulting in system gains of 5.37\% and 7.18\%, respectively. This indicates that increasing $\Gamma $ can bring better performance to the system.}
Finally, Fig. \ref{fig02d} demonstrates that the higher $T$, the better secrecy performance since more data can be collected. 

Fig. \ref{fig03} presents the 2D and 3D trajectories for different schemes. 
When the power is fixed, in order to reduce interference with $P $, the degree of freedom of $J $'s trajectory will decrease. Comparing Figs. \ref{fig03e} and \ref{fig03g}, under the UJNP scheme, $J $ will fly higher in the trajectory near $P $ to control interference, which leads to performance loss.
The optimal 2D trajectory for $U$ involves sequentially visiting each GD in order, following straight-line paths between GDs to conserve energy, as demonstrated in Figs. \ref{fig03a} and \ref{fig03d}. 
For $J$, the optimal 2D trajectory involves rapidly approaching $E$ and then maintaining close pursuit until reaching the end position, shown in Figs. \ref{fig03a} - \ref{fig03d}. 
Moreover, to achieve a favorable elevation angle, $U$ must first climb to the maximum altitude along the $z$-axis before descending toward GDs, which can be found in Fig. \ref{fig03e}. 
To effectively suppress the eavesdropper, $J$ needs to stay as close as possible to $E$ in the vertical dimension while maintaining a safe horizontal distance, given in Figs. \ref{fig03e} - \ref{fig03g}. 
Since $J$ operates at the minimum altitude, $J$ in UJ2D must be closer to $E$ than in the other schemes, by comparing Figs. \ref{fig03a} - \ref{fig03d}.

Figs. \ref{fig04a}-\ref{fig04c} plot the optimized transmit power of GDs and $J$ and Fig. \ref{fig04d} presents the SSE of the considered systems for each time slot.   
One can observe that there are four high-power regions because when $U$ is near a GD, the GD uses larger transmit power to maximize the secrecy rate, corresponding to the four high-SE regions in Fig. \ref{fig04d}. 
When $U$ flies between GDs, where channel conditions are relatively poor, the GD uses lower power. Compared to Fig. \ref{fig04b}, the variation in transmit power of the GDs in Figs. \ref{fig04a} and \ref{fig04c} is smaller. This is because, in both the proposed strategy and UJ2D, the trajectory of $U$ is optimized as needed. In contrast, in FUOJ, $U$ operates with a fixed trajectory, resulting in larger variations in the distance between $U$ and the scheduled GD and the worst SSE, as shown in Fig. \ref{fig04d}. 
As shown in the results, the SSE under the UJNP scheme is consistently lower than that under the proposed scheme in each time slot, underscoring the critical importance of trajectory design and highlighting that power control remains a key factor for performance enhancement.
Moreover, in UJ2D, $U$ can only adjust its horizontal trajectory but not its vertical trajectory, so the GDs require relatively higher transmit power. 
Additionally, since $D_1$ is close to $P$, it should utilize a relatively lower transmit power compared to the other GDs. 
Further, in the initial stage, $J$ is far from $E$ and thus uses high power to transmit AN. As it approaches $E$, lower power is sufficient to suppress eavesdropping. 
After reaching above $D_4$, since $J$ must return to the endpoint and the distance to $E$ increases, it needs to use higher transmit power again. 
Based on Figs. \ref{fig04a}-\ref{fig04c}, one also can observe that all the GDs obtain service in turn.

Fig. \ref{fig05} illustrate the (2D and 3D) trajectories and the (horizontal and vertical) velocities of $U$ with the proposed scheme with varying ${P_{{U,\mathrm{ave}}}^{{\mathrm{hor}}}}$ and ${P_{{U,\mathrm{ave}}}^{{\mathrm{ver}}}}$, respectively. 
As we know, the power budget determines whether the UAV can fly horizontally to the positions above GDs or to the altitude in the vertical direction. 
As shown in Figs. \ref{fig05a} and \ref{fig05c}, when ${P_{{U,\mathrm{ave}}}^{{\mathrm{hor}}}} = 145$, $U$ cannot approach users $D_1$ and $D_4$ horizontally, which forces it to fly higher vertically to compensate in Figs. \ref{fig05b} and \ref{fig05d} when ${P_{{U,\mathrm{ave}}}^{{\mathrm{ver}}}} = 65$. 
In contrast, when ${P_{{U,\mathrm{ave}}}^{{\mathrm{hor}}}} = 175$ and ${P_{{U,\mathrm{ave}}}^{{\mathrm{ver}}}} = 65$, $U$ can get closer to all the GDs horizontally and hover directly above them, eliminating the need to ascend to the maximum height, which is demonstrated in Figs. \ref{fig05b} and \ref{fig05d}.

Fig. \ref{fig06} shows the trajectory of $J$ and the achievable secrecy rate of the proposed scheme with varying $\Gamma$. 
It can be seen from Figs. \ref{fig06a} and \ref{fig06b} that when $\Gamma = -115$ dB, $J $ can not approach $D_1$ in horizontal and vertical directions to ensure IT restrictions on $P$.
Moreover, $J$ must use the lower power to transmit AN, shown in Fig. \ref{fig06c}. 
These reasons result in a lower SSE, shown in Fig. \ref{fig06d}.

Fig. \ref{fig07} illustrates the ASSE of different schemes with varying $T$, ${P_{{U,\mathrm{ave}}}^{{\mathrm{hor}}}}$, $r_E$, and $\Gamma$. 
The curves in Fig. \ref{fig07a} demonstrate that an increase in $T$ enhances the performance of all schemes since a larger $T$ permits $U$ to remain for a longer time over the GD. 
Fig. \ref{fig07b} testifies that a rise in ${P_{{U,\mathrm{ave}}}^{{\mathrm{hor}}}}$ results in better secrecy performance since the better channel quality between $U$ and the scheduled GD is obtained, which consequently improved the ASSE. 
In Fig. \ref{fig07c}, the effect of ${r_{E}}$ on the secrecy performance is demonstrated. We can observe that the increasing ${r_{E}}$ leads to a decreasing ASSE for all schemes since a larger ${r_{E}}$ denotes a larger estimation error of the eavesdropping channel.
Fig. \ref{fig07d} shows the influence of $\Gamma$ on the secrecy performance of the considered system. It can be observed that, as $\Gamma$ increases, the ASSE improves. However, in the larger-$\Gamma$ region, the ASSE becomes independent of $\Gamma$ since the system enters the non-cognitive mode. 

\color{black}
\section{Conclusion}
\label{Conclusions}

In this study, we examined the secrecy performance of a dual-UAV-assisted system in a data collection scenario with an aerial eavesdropper. The methodology involved improving secrecy performance by jointly considering the 3D trajectories of the dual UAVs, the transmit power and scheduling of GDs, and the jamming signal power. By jointly optimizing these variables, the ASSE was maximized.
The BCD and SCA techniques were used, and the formulated problem was decomposed into several manageable convex subproblems, which were solved efficiently using the iterative algorithm we proposed.
Numerical results confirm the convergence and effectiveness of the proposed algorithm and provide valuable insights into how parameters (such as average transmit power budgets and flight duration) affect the ASSE. 
Extending the single-antenna to a multi-antenna system and integrating beamforming with ISAC technology to collaboratively estimate the locations of aerial eavesdroppers will be part of our future research.

\end{document}